\documentclass[10pt]{article}
\usepackage{epsfig}
\usepackage{amsmath,amssymb}
\numberwithin{equation}{section}
\newcommand{\be}{\begin{equation}}
\newcommand{\ee}{\end{equation}}
\newcommand{\bea}{\begin{eqnarray}}
\newcommand{\eea}{\end{eqnarray}}

\begin{document}
\title{{\bf Phase transitions in the early and the present Universe: from the
big bang to heavy ion collisions\footnote{Lectures delivered at the
Nato Advanced Study Institute: Phase Transitions in the Early Universe:
Theory and Observations. Erice, 6th-17th December 2000. Eds, H. J. de
Vega, I. Khalatnikov, N. Sanches.}}}
\author{{\bf D. Boyanovsky}\\Department of Physics and Astronomy, University of
Pittsburgh,\\
Pittsburgh  PA. 15260, U.S.A}
\date{\today}
\maketitle
\begin{abstract}


In these lectures I discuss cosmological phase transitions with the
goal of  establishing the possibility of observational consequences. I
argue that the {\em only} phase transition amenable of experimental
study within the foreseeable future is that predicted by QCD and
discuss some of the potential observational cosmological consequences
associated with this phase transition(s). I describe the experimental
effort to study the QCD phase transition(s) at RHIC and  SPS and
summarize some of the recent experimental results. The possibility of
novel phases of QCD in the core of pulsars is discussed along with the
suggested observational consequences.  A brief review of standard big
bang cosmology as well as the astrophysics of compact stars sets the
stage for understanding the observational cosmological and
astrophysical consequences of  phase transitions in the standard model.

\end{abstract}

\tableofcontents

\section{Prologue: Cosmological Phase Transitions, Theory vs.
Observations}

The theme of this School is Cosmological Phase Transitions, Theory and
Observation.

A wealth of  cosmological data  is providing confirmation of sound
theoretical ideas in early Universe cosmology. Measurements of
temperature anisotropy in the CMB by satellite, balloon borne and earth
based observations as well as measurements of the acceleration of the
expansion with supernovae Type Ia searches and precision measurements
of light element abundances (for recent reviews see~\cite{turner})
provide an impressive body of complementary high quality data.

While the unprecedented quantity and quality of cosmological data seems
to validate the main ideas of early Universe cosmology,  the
observational consequences of cosmological {\em phase transitions} are
still rather indirect. Although  there is more theory than {\em direct}
observation of aspects of cosmological phase transitions, I will argue
in these lectures that current and future accelerator experiments along
with observations of the properties of pulsars will be opening a window
to the early Universe at a time scale of about $10^{-6}$ seconds after
the Big Bang.

To set the stage for the description of the experimental effort to
probe the {\em last} phase transition predicted by the standard model
of particle physics, and to understand its potential cosmological
impact, I  review the standard hot big bang model and the astrophysics
of compact stars, emphasizing the different scales and the observations
associated with relevant phenomena.

\section{The Standard Hot Big Bang}
\subsection{ Ingredients}
The standard ``Hot Big Bang'' theory of Early Universe cosmology
is based on a wealth of observations and rests upon the following
pillars:
\begin{itemize}
\item{Homogeneity and isotropy: on large scales $ \geq 100\mbox{Mpc}$ the Universe looks
 homogeneous and isotropic. This
is confirmed by galactic surveys of large scale structure and by
the homogeneity and isotropy of the Cosmic Microwave Background
(CMB). }

\item{ The Hubble expansion: objects that are separated by  a (comoving) distance $d$
recede from each other with a velocity $v= H~d$ with $H$ the Hubble
parameter (or Hubble constant) whose value today is $H_0 \sim 65
\mbox{Km/s/Mpc}$. The Hubble law of expansion determines the size of
the {\em causal horizon}, objects separated by a comoving distance }
\bea
d_H & = &  3000 h^{-1} \mbox{Mpc} \label{horizon}\\
h & = & \frac{H}{100 \mbox{Km/s/Mpc}} \label{littleh} \eea \noindent
recede from each other at the speed of light and are therefore causally
disconnected. In particular the size of the visible (causal) horizon
today is $\sim 3000$~Mpc.

\item{The fossil Cosmic Microwave Background (CMB) radiation: The Universe is immersed in a
bath of thermal photons
at a temperature $T_0 = 2.73 \mbox{K}$ with an almost perfect blackbody
distribution. This distribution and small anisotropies of order $\Delta
T/T_0 \sim 10^{-5}$ were measured in 1992 by the COBE satellite  and
their detection represents a triumph for the standard hot big bang
model~\cite{smoot}. The small temperature anisotropies, predicted by
cosmological models, provide the clue to the origin of galaxy formation
and large scale structure and is an important confirmation of theories
of early Universe cosmology.   }

\item{The abundance of light elements: Observations of the abundance of elements in low metallicity
 regions reveals
that about $76\%$ of matter is in the form of Hydrogen, about $24\%$
(by mass) in ${}^4He$ and very small abundances of $^3He$ ($\sim
10^{-5}$), deuterium (D) ($\sim 10^{-5}$) and $^7Li$ ($\sim 10^{-10}$)
all relative to Hydrogen. These elements were formed during the first
three minutes after the Big Bang, while heavier elements are produced
in the interior of stars and astrophysical processes such as supernovae
explosions.  }

\end{itemize}
\subsection{The building blocks:}
The main building blocks for a theory of the standard Hot Big Bang
are:
\begin{itemize}
\item{\underline{\bf Gravity}: {\em Classical} general relativity provides a good description of the geometry
of space time for distances $l\geq l_{Pl} \sim 10^{-33}\mbox{cm}$ or
time scales $t \geq t_{Pl} \sim 10^{-43}\mbox{s}$, or equivalently
energy scales smaller than the Planck scale $M_{Pl} \sim
10^{19}\mbox{Gev}$. We  have to wait for a consistent quantum theory of
gravity unified with matter  to explain phenomena on shorter space-time
scales or larger energy scales.

 Homogeneity and isotropy lead to the
Robertson-Walker metric

\begin{equation}\label{FRWmetric}
ds^2 = dt^2- a^2(t) \left[\frac{dr^2}{1-kr^2}+r^2 \left(d\theta^2
+ \sin^2(\theta)~ d\phi^2 \right) \right]
\end{equation}
\noindent where $t$ is the comoving time. The constant $k$
determines the spatial curvature and can be set to be either $\pm
1$ or $0$ by redefining the scale of coordinates. For $k=1,0,-1$
the Universe is closed, flat or open respectively. The scale
factor $a(t)$ relates physical and comoving scales

\begin{equation}\label{scale}
l_{phys}(t)= a(t) l_{com}
\end{equation}
The Friedmann equation determines the evolution of the scale
factor from the energy density
\begin{equation}\label{friedeqn}
\left(\frac{\dot{a}(t)}{a(t)} \right)^2\equiv H^2(t) = \frac{8\pi
\rho}{3 M^2_{Pl}}- \frac{k}{a^2(t)}
\end{equation}

A spatially flat Universe has the critical density

\begin{equation}\label{rhocrit}
\rho_c= \frac{3}{8\pi}M^2_{Pl}H^2
\end{equation}

\noindent and it is customary to introduce the ratio of the density of
any component (radiation, matter etc)  to the critical density as

\begin{equation}\label{bigomega}
\Omega= \frac{\rho}{\rho_c}
\end{equation}

 The energy momentum
tensor is assumed to have the fluid form leading to the (fluid)
conservation of energy equation
\begin{equation}\label{conener}
\dot{\rho}+3\left( \rho+p\right) \frac{\dot{a}}{a} =0
\end{equation}
\noindent where $\rho,p$ are the energy density and pressure
respectively. The two equations (\ref{friedeqn},\ref{conener}) can be
combined to yield the acceleration of the scale factor,
\begin{equation}\label{acceleration}
\frac{\ddot{a}}{a}= -\frac{4\pi}{3M^2_{Pl}}(\rho + 3p)
\end{equation}
\noindent which will prove useful later.
 In order to provide a close set of equations we must append an
``equation of state'' $p=p(\rho)$ which is typically written in
the form

\begin{equation}\label{eqnofstate}
p=w(\rho)\rho
\end{equation}
 Current
observations~\cite{turner} favor a spatially flat Universe $k\equiv 0$
which is consistent with predictions from inflationary scenarios.
 For a spatially flat Universe $k=0$ and for  $w={\rm constant}$   we obtain  the following
 important cases:

\begin{eqnarray}
&& w=0 : {\rm Matter ~ domination}  \Rightarrow \rho \propto
a^{-3}~;~
a(t) \propto t^{\frac{2}{3}} \label{MD} \\
&& w=\frac{1}{3} : {\rm Radiation ~ domination}  \Rightarrow \rho
\propto a^{-4}~;~
a(t) \propto t^{\frac{1}{2}} \label{RD} \\
&& w=-1 : {\rm De~Sitter~ expansion}  \Rightarrow \rho = {\rm
constant}~;~ a(t) \propto e^{Ht} ~;~H=
\sqrt{\frac{8\pi\rho}{3M^2_{Pl}}} \label{DS}
\end{eqnarray}

Furthermore we see that for accelerated expansion it must be the
case that $w < -1/3$.

 }

\item{\underline{\bf The Standard Model of Particle Physics:} the current standard model of particle physics, experimentally
tested with remarkable precision describes the theory of strong
(QCD), weak and electromagnetic interactions (EW) as a gauge
theory based on the group $SU(3)_c \otimes SU(2) \otimes U(1)$.
The particle content is: 3 generations of quarks and leptons:   }

\[
\left( \begin{array}{c} u\\d
\end{array}\right)
\left(\begin{array}{c} c\\s
\end{array}\right)
\left(\begin{array}{c} t\\b
\end{array}\right)\; ; \;
\left(\begin{array}{c} \nu_e\\e
\end{array}\right)
\left(\begin{array}{c} \nu_{\mu}\\\mu
\end{array}\right)
\left(\begin{array}{c} \nu_{\tau}\\\tau
\end{array}\right)
\]
vector Bosons: 8 gluons (massless) , $Z_0$ (mediate neutral currents),
$W^{\pm}$ (mediate charge currents) with masses of order $80,90$ Gev
respectively and the photon (massless) and scalar Higgs bosons,
although the experimental evidence for the Higgs boson is still
inconclusive.

\end{itemize}

Current theoretical ideas propose that the strong, weak and
electromagnetic interactions are unified in a grand unified quantum
field theory (GUT), perhaps with supersymmetry as the underlying
fundamental symmetry and a unification scale $M_{GUT} \sim 10^{15}
\mbox{Gev}$. Furthermore the ultimate scale at which Gravity is
eventually unified with the rest of particle physics is the Planck
scale $M_{Pl} \sim 10^{19}\mbox{Gev}$. Although there are several
proposals for the total unification of forces some of which  invoke
strings, M-theory, extra dimensions and a variety of novel and
fascinating new concepts, the experimental confirmation of any of these
ideas will not be available  soon. However, the physics of the
``standard'' model of the strong and electroweak interactions that
describes phenomena at energy scales below $\sim 100$ Gev is on solid
experimental footing.

The connection between the standard model of particle physics and
early Universe cosmology is through Einstein's equations that
couple the space-time geometry to the matter-energy content. We
argued above that at energy scales well below the Planck scale
gravity can be studied classically. However, the standard model of
particle physics is a {\em quantum field theory}, thus the
question arises: how to treat classical space-time but with
sources that are quantum fields. The answer to this question is:
semiclassically, Einstein's equations (without the cosmological
constant) are interpreted as

\begin{equation}\label{einstein}
G^{\mu \nu} = R^{\mu \nu} -\frac{1}{2} g^{\mu \nu} R = \frac{8
\pi}{M^2_{Pl}}\langle \hat{T}^{\mu \nu} \rangle
\end{equation}

\noindent with $\hat{T}^{\mu \nu}$ is the {\em operator} energy
momentum tensor and the expectation value is taken in a given
quantum state (or density matrix). A state that is compatible with
homogeneity and isotropy must be translational (and rotational)
invariant, and the expectation value of the energy momentum tensor
operator must have the fluid form $\langle T^{\mu}_{\nu} \rangle=
{\rm diag}[\rho, -p,-p,-p ]$.

Therefore, through this identification the standard model of particle
physics provides the sources for Einstein's equations. All of the
elements are now in place to understand the evolution of the early
Universe from the fundamental standard model. Einstein's equations
determine the evolution of the scale factor, the standard model
provides the energy momentum tensor.


\subsection{Energy scales:}

While a detailed description of early Universe cosmology is available
in many excellent books~\cite{bookkolb}-\cite{liddlerev}, a broad-
brush picture of the main cosmological epochs can be obtained by
focusing on the energy scales of particle, and atomic physics.

{\bf Total Unification:} Gravitational, strong and electroweak
interactions are conjectured to become unified and  described by a
single quantum theory at the Planck scale $\sim 10^{19}$Gev. While
there are currently many proposals that seek to provide such
fundamental description, these are still fairly speculative and no
experimental confirmation is yet available.

\vspace{2mm}

{\bf Grand Unification:} Strong and electroweak interactions
(perhaps with supersymmetry) are conjectured to become unified at
an energy scale $\sim 10^{15}$ Gev corresponding to a temperature
$T \sim 10^{28}K$. There are very compelling theoretical reasons
(such as the joining of the running coupling constants) that lead
to this conjecture, but there is as yet no experimental evidence
in favor of these ideas.

\vspace{2mm}

{\bf Electroweak :} Weak and electromagnetic interactions become
unified in the electroweak theory based on the gauge group
$SU(2)\otimes U(1)_Y$. The weak interactions become short ranged after
a symmetry breaking phase transition $SU(2)\otimes U(1)_Y \rightarrow
U(1)_{em}$ at an energy scale of the order of the mass of the
$Z_0,W^{\pm}$ vector bosons, $E \sim 100$Gev corresponding to a
temperature $T_{EW} \sim 100$ Gev. At temperatures $T>T_{EW}$ the
symmetry is restored and  all  vector bosons are (almost) massless
(save for plasma effects that induce screening masses). For $T<T_{EW}$
the vector bosons that mediate the weak interactions (neutral and
charged currents) $W^{\pm},Z_0$ acquire masses while the photon remains
massless. Thus $T_{EW}$ determines the temperature scale of the
electroweak phase transition in the early Universe and is the {\em
earliest} phase transition that is predicted by the standard model of
particle physics.

\vspace{2mm}

{\bf QCD:} The strong interactions have a typical energy scale
$\Lambda_{QCD} \sim 200 \mbox{Mev}$ at which the coupling constant
becomes of order one. This energy scale corresponds to a temperature
scale $T_{QCD} \sim 10^{12}$K. QCD is an asymptotically free theory,
the coupling between quarks and gluons becomes smaller at large
energies, but it diverges at the scale $\Lambda_{QCD}$. For energy
scales below $\Lambda_{QCD}$ QCD is a strongly interacting theory.This
phenomenon is interpreted in terms of a phase transition at an energy
scale $\Lambda_{QCD}\sim 200$~Mev  or $T_{QCD} \sim 10^{12}$~K. For
$T>T_{QCD}$ the relevant degrees of freedom are weakly interacting
quarks and gluons, while below are hadrons. This is the quark-hadron or
confinement-deconfinement phase transition. At about the same
temperature scale QCD has another phase transition that results in
chiral symmetry breaking (for more details see  section 4). The QCD
phase transition(s) are the {\em last} phase transition predicted by
the standard model of particle physics. The high temperature phase
above $T_{QCD}$, with almost free quarks and gluons (because the
coupling is small by asymptotic freedom) is a {\em quark-gluon plasma}
or QGP for short. Current experimental programs at CERN (SPS-LHC) and
Brookhaven National Laboratory(AGS-RHIC) are studying the QCD phase
transition via ultrarelativistic heavy ion collisions and a systematic
analysis of the data gathered at SPS-CERN during the last decade has
given an optimistic perspective of the existence of the
QGP\cite{heinz,braun} (see section 4 below).

\vspace{2mm}

{\bf Nuclear Physics:}~Low energy scales that are relevant for
cosmology are determined by the binding energy of light elements, in
particular deuterium, whose binding energy is $\sim 2$~Mev
corresponding to a temperature $T_{NS}\sim 10^{10}$~K. This is the
energy scale that determines the onset of primordial nucleosynthesis,
as described below.

\vspace{2mm}

{\bf Atomic Physics:} Another very important low energy scale relevant
for cosmology corresponds to the binding energy of hydrogen $\sim
10$~eV. This is the energy scale at which free protons and electrons
combine into neutral hydrogen. As it will be described below the
relevant
 scale is more like $\leq 0.3 $~eV corresponding to a temperature $T\sim 3 \times 10^{3}$~K.

 \vspace{1mm}

Based on the ingredients described above, a very detailed picture of
the thermal history of the Universe
emerges~\cite{bookkolb}-\cite{liddlerev}: during the first $\sim 10000$
years after the Big Bang the Universe was radiation dominated expanding
and cooling (almost) adiabatically. As a consequence the  entropy $S
\propto V(t)T^3(t)\propto V_0 \left[a(t)T(t)\right]^3$ is constant,
implying $T=T_0/a(t)$. Radiation domination, in turn, results in that
$a(t)\propto t^{\frac{1}{2}}$ and a detailed
analysis~\cite{bookkolb}-\cite{liddlerev} reveals that

\begin{equation}
 T(t) \sim \frac{10^{10}\mbox{K}}{t^{\frac{1}{2}}(sec)} \sim
\frac{1~\mbox{Mev}}{t^{\frac{1}{2}}(sec)} \label{Toft}
 \end{equation}

\section{Phase Transitions and their aftermath}
\subsection{GUT's and inflation} Current theoretical ideas of theories beyond the standard model
suggest that there could have been a phase transition at the GUT scale
$T_{GUT} \sim 10^{15}~{\rm Gev}~\sim 10^{32}~{\rm K}$.  This energy
scale is also usually associated with an important cosmological stage:
inflation, during which the acceleration of the scale factor is {\em
positive} which implies via eqn. (\ref{acceleration}) that $p<-\rho/3$.
Inflation plays a very important role in early Universe cosmology and
current observations of the power spectrum of temperature anisotropies
seem to confirm the robust features of the inflationary
proposal~\cite{turner}. While I will not attempt to review all features
of inflationary cosmology, for which the reader is referred to the
literature~\cite{bookkolb}-\cite{liddlerev}, I summarize some of the
most important concepts so as to make contact with the observable
consequences of phase transitions later.

As is mentioned above, inflation corresponds to an epoch of accelerated
expansion, i.e, with $\ddot{a}(t)>0$ which from eqn.
(\ref{acceleration}) requires that $p<-\rho/3$. Within particle physics
models this is achieved by considering the energy momentum tensor of a
scalar field, which in principle is one of the fields in the GUT. For a
scalar field the energy density and the pressure are given
by~\cite{bookkolb}-\cite{liddlerev}

\begin{eqnarray}\label{enerpres}
\rho&& = \frac{\dot{\varphi}^2}{2}+ \frac{(\nabla
\varphi)^2}{2a^2(t)}+V(\varphi) \label{rho} \\
p && =\frac{\dot{\varphi}^2}{2}- \frac{(\nabla
\varphi)^2}{2a^2(t)}-V(\varphi) \label{press}
\end{eqnarray}

\noindent where $V(\varphi)$ is the scalar potential.

Inflation results from  the generalized slow-roll
condition~\cite{bookkolb}-\cite{liddlerev}

\begin{equation}\label{slowroll}
V(\varphi) \gg \dot{\varphi}^2~,~ \frac{(\nabla
\varphi)^2}{a^2(t)}
\end{equation}

\noindent which results in that $\rho = -p \sim V(\varphi) \simeq {\rm
constant}$ leading to the De~Sitter solution eqn.(\ref{DS}) for the
scale factor. This situation can be achieved via a variety of
inflationary scenarios (old, new, chaotic, hybrid
etc,)~\cite{bookkolb}-\cite{liddlerev}. With the purpose of
establishing contact with observational consequences of phase
transitions I focus the discussion on either a first order (old
inflation) or second order (new inflation) phase
transition~\cite{bookkolb}-\cite{liddlerev} at the GUT scale.

In these situations the expectation value of the scalar field is at a
false vacuum extremum of the potential with a very slow time evolution.
The expansion of the Universe red-shifts the inhomogeneities of the
field and the slow-roll condition (\ref{slowroll}) is fulfilled. In
these scenarios the expectation value of the scalar field is nearly
constant  during the inflationary stage leading to a De~Sitter
expansion of the scale factor as in eqn. (\ref{DS}) with

\begin{equation}\label{inflaH}
H= \sqrt{\frac{8\pi V(\varphi)}{3M^2_{Pl}}}
\end{equation}

Inflation terminates when the field ``rolls down'' to the minimum of
potential, where further oscillations result in particle production,
reheating and a transition to a standard hot big bang, radiation
dominated era~\cite{bookkolb}-\cite{liddlerev}. While the evolution of
the scale factor along with the dynamics of the scalar field (inflaton)
were typically studied using the {\em classical} equations of motion
for the scalar field, more recently a consistent description of the
{\em quantum dynamics} has been
provided~\cite{noscos,inflationPT,dynamicsPT}. The non-equilibrium
dynamics of cosmological phase transitions requires a non-perturbative
framework that treats self-consistently the dynamics of the matter
field and the evolution of
 the metric. This framework leads to a detailed understanding of the
classicalization of quantum fluctuations as well as to a microscopic
justification for inflation in a full quantum field theory. We refer
the reader to ref.~\cite{noscos,inflationPT,dynamicsPT} for more
details which fall outside the main scope of these lectures.

An important aspect of inflationary dynamics and one that is very
relevant to the discussion of observables associated with phase
transitions stems from the positive acceleration of the scale factor. A
simple calculation shows that

\begin{equation}\label{physscales}
\frac{\ddot{a}(t)}{a(t)} > 0 \Longrightarrow \frac{\dot{a}(t)}{a(t)} >
\frac{\dot{d}_H(t)}{d_H(t)}
\end{equation}

Therefore during a period of accelerated expansion or inflation, {\em
the scale factor grows faster than the Hubble radius}. In particular
for De~Sitter inflation while the Hubble radius $d_H = 1/H$ is
constant, the scale factor grows exponentially $a(t) = a_0 e^{Ht}$.
This feature of accelerated expansion is indeed remarkable. Consider a
perturbation of physical wavelength $\lambda_{phys}(t)=\lambda_c a(t)$
where $\lambda_c$ is the comoving wavelength. When
$\lambda_{phys}(t)<d_H$ causal microphysical processes can affect this
perturbation, but when this wavelength ``crosses the horizon'' i.e,
when $\lambda_{phys}(t)> d_H$ no causal process can affect the
perturbation. Thus  physical wavelengths {\em inside} the Hubble radius
(or loosely speaking the causal horizon) are causally connected and
influenced by microphysical processes. When these wavelengths cross the
horizon they stretch {\em superluminally}  and ``decouple'' from causal
processes, hence their evolution is not affected by microphysical
processes.

We note that for matter or radiation dominated cosmologies $a(t)\propto
t^{2/3}~,~t^{1/2}$ respectively, for which the Hubble radius $d_H(t) =
3t/2 ~,~ 2t$ respectively,  grows {\em faster} than the scale factor.
Therefore wavelengths that at some time are outside the Hubble radius,
eventually ``cross the horizon'' back inside the Hubble radius.

Since inflation is followed by the standard hot big bang cosmology with
a radiation and matter dominated eras, physical wavelengths that are
{\em inside} the Hubble radius during inflation, and cross {\em
outside}  during inflation will re-enter during the radiation or matter
dominated eras. This situation is depicted in figure
(\ref{fig:inflation}) below.


\begin{figure}[h]
\begin{center}
\epsfig{file=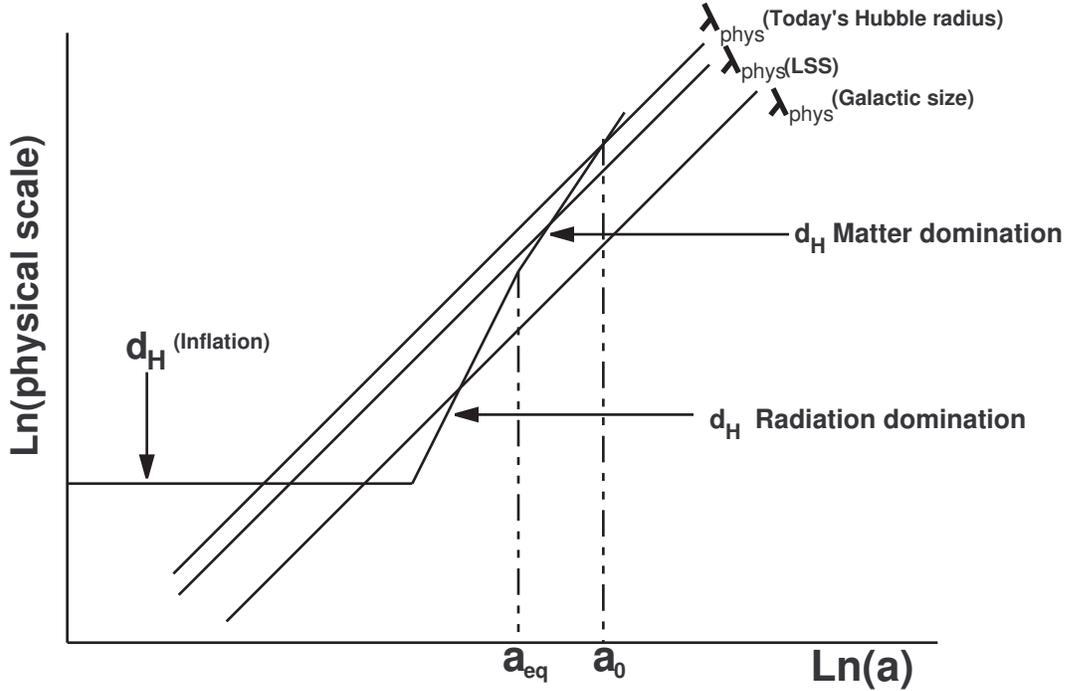} \caption{Logarithm of physical scales vs.
logarithm of the scale factor. The causal horizon $d_H$ is shown for
the inflationary (De~Sitter), radiation dominated and matter dominated
stages. The physical wavelengths ($\lambda_{phys}$) for today's Hubble
radius, the surface of last scattering (LSS) and a galactic scale are
shown. $a_{eq}~,~a_0$ refer to the scale factor at matter-radiation
equality and today, respectively. } \label{fig:inflation}
\end{center}
\end{figure}


The wavelengths that are of cosmological relevance {\em today} say with
$1~ {\rm Mpc} \lesssim  \lambda_{phys}(t_0) \lesssim 3000 ~ {\rm Mpc}$
crossed the horizon during the last $\sim 10~ {\rm e-folds}$ of
inflation\cite{bookkolb}.

 \subsubsection{Density  perturbations and the signature of a phase transition}

This important feature of accelerated expansion and inflation, i.e,
that physical wavelengths within the causal horizon during inflation
cross-out and re-enter during radiation or matter domination, provides
the mechanism for ``seeding'' temperature anisotropies. Inflation
``seeds''
 temperature inhomogeneities in the cosmic microwave background and the
matter density inhomogeneities that lead to
 large scale structure formation from primordial quantum fluctuations whose wavelengths were
 inside the Hubble radius during inflation\cite{bookkolb}-\cite{liddlerev}.

 To see this more clearly, consider small amplitude perturbations of the homogeneous scalar field
 that drive inflation

 \begin{equation}\label{pert}
 \varphi({\vec x},t)= \varphi(t)+\delta\varphi({\vec x},t)
 \end{equation}

This perturbation induces a small perturbation in the energy density,
which to linear order in $\delta\varphi({\vec x},t)$ and taking the
spatial Fourier transform is~\cite{bookkolb}-\cite{liddlerev}

\begin{equation}\label{delrho}
\delta_{\vec k} = \frac{\delta \rho_{\vec k}}{\rho} \propto \delta
\varphi_{\vec k}
\end{equation}

The power spectrum of density perturbations is obtained from the
quantum average

\begin{equation}\label{spectrum}
\langle |\delta_{\vec k}|^2 \rangle \propto \langle
|\delta\varphi_{\vec k}|^2 \rangle.
\end{equation}

An important and robust result from inflation is
that~\cite{bookkolb}-\cite{liddlerev}

\begin{equation}\label{indes}
\langle |\delta_{\vec k}|^2 \rangle \propto k^{n_s-1} ~;~ {\rm with}
~n_s \sim 1
\end{equation}
This power law spectrum with $n_s \sim 1$ is (almost) scale
invariant and referred to as Harrison-Zeldovich. As it will be
discussed in the section on the CMB below, this power spectrum is
measurable through the temperature anisotropies at the surface of
last scattering~\cite{bookliddle,bookkolb}.

The main point of this discussion is that inflation generates
temperature anisotropies and density perturbations from {\em quantum
fluctuations} whose wavelengths cross the horizon during inflation and
re-enter just before recombination~\cite{bookkolb,booklinde,bookliddle}
(see fig.~\ref{fig:inflation}).

Phase transitions during inflation modify the power spectrum of the
quantum fluctuations. After the wavelength of these fluctuations
crosses the horizon during inflation, they evolve acausally carrying
with them the information of the phase transition without being
affected by microphysical processes. When the wavelengths of these
fluctuations  re-enter the horizon near recombination the information
of the phase transition is imprinted in the temperature anisotropies in
the CMB through departures from scale invariance in the power
spectrum~(\ref{spectrum}), i.e, $n_s \neq 1$ as a consequence of the
phase transition.

To illustrate these ideas in a simple setting, let us consider a simple
scalar field theory with the potential

\begin{equation}\label{potential}
V(\varphi) = \frac{\lambda}{4} \left(\varphi^2-
\frac{m^2}{\lambda}\right)^2
\end{equation}

This is the typical potential that leads to (a second order) phase
transition. The strength of the self-coupling $\lambda$ is constrained
by the amplitude of the temperature anisotropies to be~\cite{booklinde}
$\lambda \sim 10^{-12}$. During the phase transition, the expectation
value of the scalar field $\varphi \sim 0$. Requiring that the energy
density is given by the GUT scale

$$\rho \sim \frac{m^4}{4\lambda}  \sim (10^{15}~{\rm Gev})^4$$

\noindent during the epoch of the phase transition determines that
$m \sim 10^{12}~{\rm Gev} $. The Hubble constant is then given
during this stage by

\begin{equation}\label{hubble}
H^2 \approx \frac{2\pi}{3} \frac{m^4}{\lambda M^2_{Pl}}~
\Longrightarrow ~\frac{H}{m} \sim {\cal O}(1)
\end{equation}

Therefore during the phase transition when $\varphi \sim 0$ the
expansion is of De~Sitter type and small amplitude fluctuations of the
scalar field (in terms of the spatial Fourier transform)  $\delta
\varphi_{\vec k}$ obey the linearized equation

\begin{equation}\label{fluceqns}
\ddot{\delta \varphi}_{\vec k}(t) + 3 H \dot{\delta \varphi}_{\vec
k}(t)+ \left[\frac{{\vec k}^2}{a^2(t)}- m^2\right] \delta \varphi_{\vec
k}(t)=0.
\end{equation}

We note that the sign of the mass squared term is negative as required
to describe a symmetry breaking phase transition. Setting $H=0~,~a =1$,
i.e, in Minkowsky space time we recognize that long-wavelength
fluctuations with $k^2<m^2$ grow exponentially. These are the {\em
spinodal instabilities}, a hallmark of the process of phase separation
during the phase transition~\cite{boylee,dynamicsPT}. During the stage
of De~Sitter inflation, eqn.(\ref{fluceqns}) has the following
solution~\cite{inflationPT}
\begin{eqnarray}\label{soldesitter}
\delta \varphi_{\vec k}(t) & = &  e^{-\frac{3Ht}{2}}\left[A_k
J_{\nu}\left( \frac{k}{H}e^{-Ht}\right)+ B_k J_{-\nu}\left(
\frac{k}{H}e^{-Ht}\right) \right] \nonumber \\
\nu & = & \sqrt{\frac{9}{4}+\frac{m^2}{H^2}}
\end{eqnarray}

\noindent the coefficients $A_k,B_k$  are determined by initial
conditions\cite{inflationPT}. When the physical wavevector crosses the
horizon, i.e, when $k~e^{-Ht} \ll H$  the solution grows exponentially
and is given by

\begin{equation}\label{growin}
\delta \varphi_{\vec k}(t) \propto e^{(\nu -\frac{3}{2})Ht}
\end{equation}

In the limit $H<<m$ this solution displays the spinodal long-wavelength
instabilities $\propto e^{mt}$ associated with the phase transition in
Minkowski space-time~\cite{boylee,dynamicsPT}. Thus the exponential
growth of superhorizon fluctuations given by (\ref{growin}) is a direct
consequence of the spinodal instabilities associated with the phase
transition {\em during the De~Sitter inflationary epoch}. A detailed
calculation of the power spectrum of density
fluctuations~\cite{inflationPT,guthpi} yields

\begin{equation}\label{powerspec}
\langle | \delta_{\vec k}|^2 \rangle \propto
k^{-2(\nu-\frac{3}{2})} \Rightarrow n_s = 1-2(\nu-\frac{3}{2})
\end{equation}

\noindent since $\nu > 3/2$ we see that there is more power at long
wavelengths. This is obviously a consequence of the spinodal
instabilities associated with the phase transition. Thus we arrive at
an important result that provides an observational signal of the phase
transition: {\em the power spectrum of density perturbations is tilted
to the red, i.e, the index is smaller than one. This is because the
spinodal instabilities associated with the phase transition imply
larger amplitudes for longer wavelengths, thus enhancing the infrared}.
After these perturbations re-enter the horizon close to recombination,
this power spectrum is imprinted in the temperature anisotropies of the
CMB (see the discussion on the CMB below) which are measured.  The four
year COBE-DMR Sky Map\cite{cobe4year} gives $n_s \sim 1.2 \pm 0.3$,
thus a (second order) phase transition at the GUT scale is consistent
with this power spectrum if $H/m \geq 3$ which, in turn,  is consistent
with GUT scale inflation as can be seen from eqn.(\ref{hubble}).

This simple example thus provides a good idea of the potential
observational signatures of phase transitions at the GUT scale
associated with an inflationary era: a power spectrum with a red-tilt
results  if the phase transition is triggered by spinodal instabilities
during the inflationary epoch.

Another potential observable from an inflationary phase transition was
proposed recently~\cite{sarkar}. These authors pointed out that the
breaking of scale invariance in the primordial power spectrum by an
inflationary phase transition could lead to a step-like spectral
feature, which seems to be compatible with recent measurements of the
CMB and large scale surveys (APM).

Thus observations of the temperature anisotropies in the CMB could
provide information on inflationary phase transitions. Furthermore, a
``reconstruction'' program  seeks to extract some aspects of the scalar
potential from the CMB temperature anisotropies~\cite{reconstruction},
therefore providing a further link between  phase transitions during
the inflationary era and the temperature anisotropies of the CMB.

\subsection{The Electroweak Phase Transition: Baryogenesis?}

The EWPT  is the {\em first} symmetry breaking phase transition
predicted by the standard model of particle physics. It occured at a
temperature $T_{EW} \sim 100 ~{\rm Gev}\sim 10^{15}{\rm K}$ at about
$t\sim 10^{-12}~ {\rm secs}$ after the Big  Bang when the Hubble radius
was $d_H \sim 10^{-1} {\rm cm}$. The symmetry breaking pattern is
$SU(2)\otimes U(1)_Y \rightarrow U(1)_{em}$. Probably the most
tantalizing observable from the electroweak phase transition {\em could
be} the {\em baryon asymmetry}. There is an asymmetry between particles
and antiparticles in the observed Universe, large regions of antimatter
would result in particle annihilations leading to a diffuse
$\gamma$-ray background and distortions of the CMB. None of which is
observed, leading to the conclusion that the Universe is made up of
particles up to the Hubble radius~\cite{trodden}. Furthermore Big Bang
nucleosynthesis provides accurate predictions for the abundance of
light elements up to $^7$Li in terms of the ratio of the baryon density
to the photon density
\begin{equation}\label{baryo}
\eta = \frac{n_b- \bar{n}_b}{n_{\gamma}}
\end{equation}
Observations of the abundance of light elements  constrain this
parameter to be in the range~\cite{bookkolb}

\begin{equation}\label{etacons}
4\times 10^{-10} \lesssim \eta \lesssim 7 \times 10^{-10}
\end{equation}

The important question is: what is the origin of the baryon asymmetry
(baryogenesis)? (for a recent review on baryogenesis
see~\cite{trodden}).

The necessary conditions required for  baryogenesis were identified
originally by Sakharov~\cite{sakharov}

\begin{itemize}
\item C and CP violation
  \item Baryon number violation
  \item Departure from thermal equilibrium
\end{itemize}

The EW theory violates maximally P (parity) and also C (charge
conjugation). CP violation is observed in the $K^0 ,\bar{K}^0$ system
(and currently also in the $B^0,\bar{B}^0$ system)  and is
experimentally determined by the parameters $\epsilon \sim 2.3 \times
10^{-3},\epsilon' \sim 5.2 \times 10^{-6}$ that measure indirect
(through mixing of eigenstates) and direct CP violation. CP violation
in the standard model is a consequence of the phase in the CKM mass
matrix for three generations.

 Baryon number violation in the EW theory
is a consequence of non-perturbative gauge (and Higgs) field
configurations that interpolate between topologically different vacua.
A transition between two adjacent vacua leads to one unit of baryon
number violation per family (for details see~\cite{trodden}). While at
zero temperature this is a vacuum tunneling process which is suppressed
by the barrier penetration factor $e^{-\frac{2\pi}{\alpha_w}}$ with
$\alpha_w \sim 1/30$,  at finite temperature the transition is
overbarrier, i.e, thermal activation. The overbarrier transitions are
unsuppressed at temperatures of the order of $T_{EW}$~\cite{trodden}.

The requirements of baryon number violation and CP violation are clear:
to generate any asymmetry the theory must allow for processes that
produce the quantum number in question (B) and to also distinguish
particle from antiparticle (CP). If the relevant processes are in
equilibrium then the rate for particles and that of antiparticles are
related by detailed balance and since CPT invariance implies that the
masses of particles are equal to those of antiparticles, the
distribution functions for both are identical  and no asymmetry can
result~\cite{bookkolb,trodden}.

The expansion of the Universe alone cannot provide the non-equilibrium
ingredient. To see this let us compare the reaction rate of weak
interactions (the electromagnetic interactions are much faster than the
weak interactions) to the expansion rate of the Universe. Since the
EWPT  takes place during a radiation dominated era with $\rho \propto
T^4$ the expansion rate is given by

\begin{equation}\label{HEW}
H \propto \frac{T^2}{M_{Pl}}
\end{equation}

The reaction rate is given by $\Gamma=n\sigma_w$ with $n$ the particle
density and $\sigma_w$ the cross section.  The typical weak interaction
cross section is $\sigma_{w} \sim G^2_F E^2$ with $G_F \sim
10^{-5}/{\rm Gev}^2$, at temperature $T$ the typical energy is $E\sim
T$ and a particle density $n \sim T^3$ . Thus the weak interaction rate
$\Gamma = n\sigma_w$ is of order

\begin{equation}\label{gammawi}
\Gamma \sim G^2_F T^5
\end{equation}
\noindent therefore the ratio

\begin{equation}\label{ratio}
\frac{\Gamma}{H} \sim G^2_F M_{Pl}T^3 \sim \left[10^3 T({\rm
Gev})\right]^3
\end{equation}

\noindent determines whether the weak interactions are in local thermal
equilibrium (LTE).  At $T \sim T_{EW}$ the ratio is $\Gamma/H \sim
10^{15}$, therefore weak interaction processes are in LTE (and
obviously electromagnetic processes will be more so).

Instead departures from equilibrium  results from a {\em strong} first
order phase transition that occurs via the nucleation of bubbles of the
true phase in a background of the metastable phase~\cite{trodden}.

Detailed numerical simulations of the EWPT\cite{kajantie} reveal that
it is first order if the Higgs mass $M_H < 80 {\rm Gev}$ but above this
value it is a smooth crossover. Recently, however, a summary of all the
data collected by the four detectors at LEP up to energy 202 Gev, seem
to indicate that the Higgs has a mass $M_H \gtrsim 115 {\rm
Gev}$\cite{higgs}. Thus it is very likely that the (minimal) standard
model cannot accomodate a strong first order phase transition that will
produce the non-equilibrium conditions sufficient for baryogenesis.
Furthermore, it is now clear that the CP violation in the (minimal)
standard model, encoded in the parameters $\epsilon,\epsilon'$ (or
alternative in the phase of the CKM matrix) is too small to explain the
observed baryon asymmetry.

Thus, the current theoretical understanding seems to suggest that the
baryon asymmetry cannot be explained by the minimal standard model
although  the Minimal Supersymmetric Standard Model (MSSM) may
accomodate all of the necessary ingredients~\cite{trodden}. Hence the
cosmological consequences of the EWPT are not very clear. Furthermore,
while accelerator experiments (Tevatron and LHC)  will probably lead to
an assessment of the Higgs mass and CP violating parameters, a {\em
direct} study of the EWPT will not be feasible soon. Indeed in order to
study it an energy {\em density} of order $\rho_{EW}\sim T^4_{EW}$ must
be achieved, but with $T_{EW}\sim 100~ {\rm Gev}$ this energy density
is $\rho_{EW} \sim 10^{11}\rho_n$ where $\rho_n \sim 0.15~ {\rm
Gev}/{\rm fm}^3$ is the energy density of nuclear matter!!.

\subsection{The QCD phase transition(s)}

The next and {\em last} phase transition predicted by the standard
model of particle physics takes place at the QCD scale $T \sim 100-200
{\rm Mev}$. It occurred when the Universe was about $t\sim
10^{-5}-10^{-6}$ seconds old and the Hubble radius was $d_H \sim 10
{\rm Km}$. QCD is an asymptotically free theory, that is the gauge
coupling constant between quarks and gluons varies with energy through
``vacuum polarization'' effects as

\begin{equation}\label{alfas}
\alpha_s(E) = \frac{4 \pi}{(11-\frac{2}{3}N_f)
\ln\left[\frac{E^2}{\Lambda^2_{QCD}} \right] }  ~~; ~~ \Lambda_{QCD}
\sim 200 {\rm Mev}~~;~~ \alpha_s = \frac{g^2}{4\pi}
\end{equation}

\noindent with $N_f$ being the number of flavors and $g$ is the
quark-gluon coupling. For $E \gg \Lambda_{QCD}$ quarks and gluons are
weakly coupled, but for $E \sim \Lambda_{QCD}$ the coupling becomes
strong, diverging at $E= \Lambda_{QCD}$. This divergence is interpreted
as a transition between a state of almost free quarks and gluons to a
state in which quarks and gluons are confined inside hadrons. This is
the {\em confinement-deconfinement transition} and the QCD scale
$\Lambda_{QCD}$ determines the range of temperature for this
transition. The high temperature phase in which quarks and gluons are
almost free is referred to as the {\em Quark Gluon Plasma} or QGP for
short.

 Because the up and down quark masses are so much smaller than
$\Lambda_{QCD}$ ($m_u \sim 5~ {\rm Mev};m_d \sim 10~{\rm Mev}$) the low
energy sector of QCD has an $SU(2)_R\otimes SU(2)_L$ symmetry in the
limit of massless up and down quarks. This chiral symmetry corresponds
to independent chiral rotations of the right and left handed components
of the quark fields. However the ground state of QCD spontaneously
breaks this chiral symmetry down to $SU(2)_{L+R}$ leading to three
massless Goldstone bosons. The small (and different) masses of the up
and down quark provide a small explicit symmetry breaking term in the
Lagrangian and as a consequence the (would be) Goldstone bosons acquire
a mass. This triplet of almost Goldstone bosons are the neutral and
charged pions, which are the lightest pseudoscalar particles with
$m_{\pi} \sim 140 ~{\rm Mev}$. Lattice simulations~\cite{karsch}
suggest that the confinement-deconfinement and chiral phase transitions
occur at about the same temperature $T \sim 160 ~{\rm Mev}$ (for a
review see~\cite{meyer}). At this temperature only the lightest quark
flavors influence the thermodynamics, the up, down and strange quarks
with $m_s \sim 150 ~{\rm Mev}$. The chiral phase transition associated
with the two lightest quark flavors is second order for massless quarks
but either slightly first order or a crossover when the mass of the up
and down quarks is accounted for, and is in the universality class of
the $O(4)$ Heisenberg ferromagnet. Including the strange quark the
situation is more complicated and depends on whether the strange quark
can be considered heavy or light, the transition becomes first order if
the strange quark is light~\cite{karsch}. The complication arises
because $m_s \sim \Lambda_{QCD}$ and is therefore neither light nor
heavy on the QCD scale.

The confinement-deconfinement transition for three flavors is likely to
be of first order~\cite{karsch,meyer} and therefore occurs via the
nucleation of hadronic bubbles in the background of a QGP. An estimate
of whether the expansion of the Universe at the QCD phase transition
results in non-equilibrium effects can be obtained in the same manner
as in the case of the EW transition, i.e,  by estimating the ratio
$\Gamma/H$ with $\Gamma$ a typical strong interaction reaction rate. A
typical cross section is $\sigma \sim  0.1-1~{\rm fm}^2$ and the number
of particles at $T_{QCD}\sim 200~{\rm Mev}$ is $n \sim T^3_{QCD} \sim
1/{\rm fm}^3$ this leads to $\Gamma \sim 1/{\rm fm}/c \sim 10^{22}~{\rm
sec}^{-1}$ and since $H \sim 10^6 ~{\rm sec}^{-1}$, therefore $\Gamma/H
\sim 10^{16}$ and again the QGP is in LTE (local thermal equilibrium).
However, just as in the case of the EWPT, non-equilibrium effects may
arise from a first order phase transition through supercooling and the
nucleation of bubbles. In this case, hadronic bubbles nucleate in the
host of the QGP releasing latent heat. After a short period of
reheating, the transformation between the QGP and the hadronic phases
occurs in LTE until all of the QGP transforms into hadrons. Most of the
hadrons decay on short time scales,  for example charged pions decay on
time scales $\sim 10^{-8}$ secs, and neutral pions on much shorter time
scale $10^{-16}$ secs. Therefore most hadrons have decayed  after a few
$\mu$secs after the QCD PT but for neutrons, with a lifetime $\sim 900$
secs and protons (with a lower limit for the lifetime of $\sim 10^{32}$
years).

 Since the QCD PT is the last phase
transition, there is an important effort to understand its potential
cosmological consequences. While the details are complex, not
completely understood and cannot be given in this short review, the
following important consequences are possible (for a recent review see
also~\cite{cosmicPTs,kapqgpuni})

\begin{itemize}
  \item {\bf Baryon inhomogeneities affect
  nucleosynthesis:}~\cite{apple} The nucleation of hadronic
  bubbles in a host of QGP can lead to inhomogeneities in the
  baryon number density, with the scale for inhomogeneity
  determined by the typical distance between bubbles.
These inhomogeneities  may result in inhomogeneous neutron
  to proton ratios which in turn can lead
  to inhomogeneous nucleosynthesis (see the section on
  nucleosynthesis below) and modify the abundance of light elements.
  For the inhomogeneities produced by bubble nucleation to modify
  the neutron to proton ratio for nucleosynthesis it must be that
  the typical separation between nucleating bubbles must be much
  larger that the proton and neutron diffusion lengths. Proton and
  neutrons have {\em different} diffusion lengths because of the
  Coulomb scattering of protons.
   An important criterion for
  inhomogeneous nucleosynthesis is that the mean scale for baryon
  inhomogeneity at the time of the QCD PT, i.e, the mean distance between nucleating bubbles
   must be of order  $\sim   1$   meter~\cite{apple}. This distance depends on the latent
  heat released during the (first order) phase transition, the
  free energy difference between the QGP and the hadron gas phases
  and the surface tension for the nucleating bubble. More recent
  analysis~\cite{christ} seems to point out that the distance between bubbles produced during homogeneous
  nucleation is $d_{nuc} \sim 1 {\rm cm}$. Larger values of $d_{nucl}$
  can arise from inhomogeneous nucleation seeded by
  ``impurities''~\cite{christ}. These inhomogeneities can be produced
  by primordial black holes (see below) as proposed by~\cite{christ} or
  by primordial density fluctuations as suggested by~\cite{ignatius}.

  However as recognized
  in the literature~\cite{apple,christ,ignatius} much more needs to be
  understood about the phase structure of QCD before a
  quantitatively reliable statement can be made. Of particular
  importance are
  details of the equation of state (Eos), surface tension, which determines the size of the
  nucleating bubbles,  and also
  the (in medium) mass of the strange quark. For a more recent
  estimate on inhomogeneous nucleosynthesis arising from a first order
  QCD PT see~\cite{jedaII,kapqgpuni}.

  \item {\bf Primordial Black Holes:}~\cite{jedamzik,schmid} An
  important aspect of a first order phase transition (with only
  one globally conserved quantity, such as baryon number) is that
  during the coexistence or mixed phase of QGP and hadron gas, the
  Gibbs construction determines that the phase transition occurs at
  constant pressure (as well as temperature). If the pressure is
  constant, then the (adiabatic) speed of sound vanishes and there
  are no restoring pressure waves. This in turn means that there
  is no restoring force to counterbalance the gravitational
  collapse and long-wavelength density perturbations will grow
  under self-gravity causing the gravitational collapse of the
  mass contained in a volume of radius given by this wavelength.
   This observation  leads to the possibility of
  formation of {\em primordial black holes}~\cite{jedamzik}(for a discussion of
  gravitational collapse and Jeans instability see section 3.6). An
  important question is what is the mass of the black hole?, to
  obtain a qualitative estimate we can calculate the mass in the
  Hubble radius at the time of the QCD PT: $M_H = 4\pi \rho~ d^3_H/3$
  with $d_H \sim 10 {\rm Km}$ the Hubble radius at the time of the
  PT. Using that $H^2 = 1/d^2_H = 8\pi \rho~/3M^2_{Pl}$ therefore
  $M_H = M^2_{Pl}d_H/2 \sim 10^{57} {\rm Gev} \sim 1 M_{\odot}$.
  Thus this order-of-magnitude estimate suggests that primordial
  black holes with mass up to $\sim 1 M_{\odot}$ may form during
  the QCD PT~\cite{jedamzik}. It is important to mention at this stage,
  that the Hubble radius at the QCD PT $\sim 10^4~m$ extrapolated to today is $\sim 0.2~{\rm pc}$ which is
  {\em   much smaller} than the Hubble radius today $\sim 3000$~Mpc.
  Therefore a large number of primordial black holes
produced during the QCD PT would be present today.

  A more detailed analysis of the
  possibility of primordial black hole formation and the ensuing
  density fluctuations has been provided recently~\cite{schmid}.
  The results of this reference, based on approximate or lattice
  EoS seem to lead to a less likely scenario for solar mass
  primordial black hole formation, but to an enhancement in the
  clumping of cold dark matter (CDM). It is fair to say, however,
  that there are still large uncertainties in the EoS and the
  relevant quantities that enter in the calculation. The
  experimental programs that seek to study the QGP in heavy ion
  colliders that are described below will lead to a more reliable
  understanding of the relevant aspects of the PT, EoS and QCD
  parameters. For more detailed aspects of primordial black holes,
  see~\cite{kaphole}.

  \item {\bf Strange Quark Nuggets:}~\cite{madsen}
  Witten
  suggested that during a first order QCD PT in the early Universe a large amount
   of strange matter could be produced~\cite{witten}.
 He also argued that strange matter could be
  absolutely stable with an energy per baryon which is less than
  the maximum binding energy (of Fe) $930{\rm Mev}$. While this
  scenario has been criticized (see Applegate and Hogan
  in~\cite{apple}), the uncertainties in the knowledge of
  the QCD EoS, the (in medium) strange quark mass and other
  relevant QCD parameters leave enough room for the possibility of
  formation of strange quark ``nuggets''~\cite{madsen}. Some of
  the consequences of the formation of strange quark nuggets
  during the (first order) QCD PT had been investigated
  in~\cite{madsen,alcock}. These strange nuggets can be part of the cold dark
  matter, and if they do not evaporate during $\sim 1$ sec after
  the QCD PT, their presence can affect nucleosynthesis
   by modifying the neutron to proton ratio, thus
  modifying the ${}^4He$ (and heavier elements) abundance~\cite{madsen}.
\end{itemize}

The different possible signatures of the QCD PT in the early Universe
as described by the scenarios above rely on the particular details of
the phase transition as well as thermodynamic parameters that cannot be
calculated in perturbation theory. Lattice gauge theory provides a
non-perturbative approach to studying many of these aspects and
although progress has been made in the field~\cite{karsch} the
complications associated with light quarks beyond the quenched
approximation still need further understanding. It is clear that
probably the best assessment of the QCD phase diagram would be obtained
from  experiments that can probe the hot and dense phase(s) of QCD.

The heavy ion programs at the CERN-SPS and the BNL-AGS, and the current
program at BNL-RHIC have as major goals to reveal the new state of
matter described by the QGP. Almost two decades of experiments at CERN
and BNL have provided a wealth of data that together reveal that this
new state of matter, predicted by QCD may have been formed in
ultrarelativistic heavy ion collisions. The recent announcements from
CERN~\cite{heinz,braun} yield convincing arguments that there is
already experimental evidence for the QGP. Thus there is the very
tantalizing possibility that the next generation of ultrarelativistic
heavy ion collisions at RHIC  and eventually the ALICE (A Large Ion
Collider Experiment) program at LHC will provide a more detailed
understanding of the QCD PT, and from that data we can learn about the
observable consequences of the same PT in the early Universe. This
program, along with a brief summary of  the recent results will be
described  in more detail in section 4.

\subsection{Nucleosynthesis}

The next stage down the ladder of  energy scales corresponds to $\sim
1~{\rm Mev}$. This is the scale of binding energy of deuterium and
determines the onset of primordial nucleosynthesis. As mentioned in the
previous section, after the QCD confinement-deconfinement phase
transition, the thermodynamics is described by a hadron gas. After
about a $\mu{\rm sec}$ after this PT most hadrons decay and only
neutrons and protons remain. Therefore the Universe can be
characterized by a plasma of $n,p, e^{\pm},\nu's,\gamma's$. Equation
(\ref{ratio}) for the ratio of a typical weak interaction reaction rate
and the expansion rate of the Universe, determines that at a
temperature $T \sim 1 ~{\rm Mev}$ the weak interactions {\em
freeze-out} (decouple), i.e, the rate for weak interaction processes is
smaller than that of the expansion of the Universe. Eqn. (\ref{ratio})
also means that the mean free path for the weak interactions is larger
than the Hubble radius for $T < 1 ~{\rm Mev}$.  A more detailed
calculation~\cite{nucleosynthesis} shows that the freeze-out or
decoupling temperature is actually $T_f=0.8~{\rm Mev}$. For $T\gg T_f$
neutrons, protons, electrons and neutrinos are in nuclear statistical
equilibrium via the weak interaction reactions

\begin{equation}\label{reactions}
n \leftrightarrow p+ e^-+\bar{\nu}_e ~;~ \nu_e+n \leftrightarrow p+e^-
~;~ e^++n \leftrightarrow p +\bar{\nu}_e
\end{equation}

Since neutrons and protons are non-relativistic near the freeze-out
temperature, their phase space distribution is determined by  the usual
Boltzmann distribution function and at any given temperature much
smaller than their masses the neutron to proton ratio is given by

\begin{equation}\label{npratio}
\frac{N_n}{N_p} = e^{-\frac{\Delta m}{T}} = e^{-\frac{1.29}{T({\rm
Mev})}}
\end{equation}
\noindent with $\Delta m$ the neutron-proton mass difference.
 The neutron to proton ratio at $T \sim
T_f$ is given by

\begin{equation}\label{npratiofrozen}
\frac{N_n}{N_p} = e^{-1.29/0.8}\sim 1/6
\end{equation}

After the Universe has cooled to the freeze-out temperature $T_f$
this ratio remains fixed but for the decay of the neutron with a
lifetime $\sim 900$ secs.

 Nucleosynthesis begins via the
reaction
\begin{equation}\label{nucleo1}
n+p \leftrightarrow d+\gamma
\end{equation}

and continues through the following two possible paths towards the
conversion of all neutrons into $^4He$

\begin{eqnarray}\label{nucleo2}
d+d \leftrightarrow  {}^3He+n    & ~~;~~  or & d+d \leftrightarrow{}^3H+p \nonumber \\
{}^3He+d \leftrightarrow {}^4He+p  &  ~~;~~ or &  {}^3H+d
\leftrightarrow {}^4He+n
\end{eqnarray}

The binding energy of deuterium is $2.22 ~{\rm Mev}$ and {\em a priori}
one would think that once the temperature falls below this value the
reaction (\ref{nucleo1}) will mainly go to the right. However, because
the number of photons is so much larger than the number of baryons
since $n_B/n_{\gamma}\propto \eta$ with $\eta$ given by eqn.
(\ref{etacons}) {\em there are too many photons in the high energy tail
of the blackbody distribution} to photodisintegrate any deuterium
formed until the temperature actually falls to $T\sim 0.1 ~{\rm Mev}$
at a time of about 2 minutes~\cite{bookkolb}. This is the ``deuterium
bottleneck'': nucleosynthesis does not begin in earnest until the first
step (\ref{nucleo1}) happens. The formation of deuterium is hindered by
the fact that the Saha equation for the equilibrium abundance for a
species of atomic number A, contains a factor $\propto
\eta^{A-1}$~\cite{bookkolb}. From the time of freeze-out when the
temperature was $T\sim T_f$ until deuterium is formed, the neutrons
have decayed via the weak interactions (with a lifetime $\sim 900$
secs) to a neutron to proton ratio $N_n/N_p \sim
1/7$~\cite{bookkolb,nucleosynthesis}. As soon as deuterium is formed
the second stage(s) (\ref{nucleo2}) occur fairly
fast~\cite{bookkolb,nucleosynthesis} and all neutrons are used up in
the formation of ${}^4He$.  The neutron to proton ratio determines the
abundance of ${}^4He$: each atom of ${}^4He$ has two protons and two
neutrons, but for each two neutrons 14 protons are needed because of
the ratio $1/7$. Altogether there are 16 baryons, out of which one
Helium atom was formed, so the abundance of Helium is $25\%$!. A
network of nuclear reactions leads to the formation of light elements
up to ${}^7Li$~\cite{nucleosynthesis}. However because there is no
nucleus with $A=5$ and no stable one with $A=8$ with unstable nuclei in
between ($^6Li,^7Be$) only small quantities of light elements other
than $^4He$ are produced. An important ingredient in these
calculations, besides the nuclear cross sections that determine the
nuclear reaction rates that have been measured in low energy
experiments is the baryon to entropy ratio $\eta$ which in turn can be
related simply to $\Omega_bh^2$ where $\Omega_b$ is the ratio of the
baryon density to the critical density~\cite{bookkolb,nucleosynthesis}
$\Omega_bh^2 \sim 3.64 \times 10^7 \eta$.

Big Bang nucleosynthesis provides a precise estimate of the baryon
density of the Universe by comparing to observations of the abundance
of deuterium. The most recent observations~\cite{deuterium} constrain
the baryon density inferred from {\em standard} big bang
nucleosynthesis to the interval

\begin{equation}\label{omegab}
0.015 \lesssim \Omega_b h^2 \lesssim 0.024
\end{equation}

\noindent with $\Omega_b$ being the baryon density expressed as a
fraction of the critical density and $h$ has been introduced in eqn.
(\ref{littleh}). In the past year, however, combined results on the CMB
anisotropies from the balloon experiments BOOMERANG and MAXIMA and
observations from type Ia supernovae have set new limits for the baryon
density~\cite{boom} (independent from the observation of abundances)
\begin{equation}\label{newomegab}
0.022 \lesssim \Omega_b h^2 \lesssim 0.039
\end{equation}

Although these results may be preliminary and require other
observations for definite
 confirmation, taken together they signal the possibility of
non-standard nucleosynthesis. These new observational constraints
on $\Omega_b$ have rekindled the interest on the possibility of
inhomogeneous nucleosynthesis~\cite{jedaII} with encouraging
results for the abundances of deuterium and helium but the
abundance predicted by these models of inhomogeneous
nucleosynthesis for heavier metals is still well below the
observed values.

The important point in this discussion and the main reason for delving
into the subject of nucleosynthesis is to highlight two important
factors that can be influenced by the QCD phase transition: i) the
helium abundance depends crucially on the neutron to proton ratio,
although it is only slightly sensitive to the baryon to photon ratio.
ii) The abundance of heavier elements up to ${}^7Li$ depends in an
important manner on the baryon to photon ratio. The point being made
here is that if the QCD phase transition is of first order, there is
the potential for producing inhomogeneities that  modify the neutron to
proton ratio, hence the helium abundance, as well as the baryon to
photon ratio, and therefore the abundance of heavier elements. Hence if
the value of $\Omega_bh^2$ favored by BOOMERANG-MAXIMA and Type Ia
supernovae searches stand up to further and deeper scrutiny, and are
still  in discrepancy  with standard big bang nucleosynthesis perhaps
inhomogeneous nucleosynthesis at the QCD phase transition could provide
an explanation or be a part of the explanation.

\subsection{Recombination, LSS and CMB}

Continuing down the ladder of energy scales, the next step corresponds
to energies of order eV determined by the binding energy of the
hydrogen atom. Actually two different (but related) processes occur at
this energy scale: a) (re) combination of electrons and protons into
neutral hydrogen atoms which is described by the chemical reaction

\begin{equation}\label{recombo}
e^-+p \leftrightarrow H+\gamma
\end{equation}

\noindent b) photon decoupling (or freeze out): electrons and
photons are in LTE through Thompson scattering $e^-+\gamma
\rightarrow e^-+\gamma$. This electromagnetic process has the
(large) cross section

\begin{equation}\label{thompson}
\sigma_T = \frac{8 \pi \alpha^2}{3 m^2_e} = 6.6 \times
10^{-25}{\rm cm}^2
\end{equation}

When the reaction rate $\Gamma_{\gamma} = n_e \sigma_T$ with $n_e$ the
electron density becomes smaller than the expansion rate (or
alternatively the photon mean free path becomes larger than the Hubble
radius) the photons do not scatter any more, their distribution freezes
and redshifts with the expansion. When the free electrons combine with
protons binding into neutral hydrogen atoms the reaction rate becomes
very small. Detailed calculations~\cite{bookkolb} reveal that
recombination and decoupling occur at a temperature $T_R \sim 3000 K$
at a redshift $z\sim 1100$. The reason that the recombination
temperature is {\em not} given by the binding energy of the hydrogen
atom ($=13.6~{\rm eV}$) is again  a consequence of the fact that the
baryon to photon ratio is so small as given by (\ref{etacons}) that
even a small number of energetic photons in the blackbody tail can
photodisintegrate  the hydrogen atoms formed at $T>T_R$.

The important point here is that at a redshift $z\sim 1100$ and
temperature $T_R \sim 3000~ K$ photons freeze-out, i.e, they do not
undergo any further scattering. This occurs at a time $t_R \sim 300000$
years after the Big Bang which  defines the Last Scattering Surface
(LSS). Photons have been travelling to us from the last scattering
surface  without scattering for the last $\sim 15 $ billion years and
carry information of the LSS.

The discovery of the temperature anisotropies in the CMB by the
COBE satellite with $\Delta T/T \sim 10^{-5}$ on angular scales
from $90^o$ down to about $2^o$ raises {\em two} fundamental
questions:

\begin{itemize}
  \item {\bf The Horizon Problem:} photon decoupling occurs at
  $t\sim 300000$ years after the Big Bang at a redshift of $z\sim 1100$. At this time  the Hubble radius
  is $d_H(t_R) \sim 100~{\rm Kpc}$ which determines the size of a correlated region.
   {\em Today} the size of this correlated patch  is  $\sim (1+z)d_H(t_R) \sim 100~ {\rm Mpc}$, however the
  Hubble radius today is $\sim 3000~{\rm Mpc}$. Therefore we
  would expect that in the Hubble volume today there are about $\sim
  30000$ of these regions. Inside each of these regions the temperature fluctuations would be
 correlated but the regions would be completely uncorrelated
  between them. Each correlated patch today subtends an angle of about $2^o$ which is
  the angular scale of the LSS today.
   However the CMB is homogeneous to 1 part in $10^5$ on angular
   scales between $2^o$ up to $90^o$ as revealed by COBE. Thus the
   question: how did these regions that were outside the causal
   horizon of each other at photon decoupling manage to establish correlations and be
   so homogeneous in temperature?.

  \item {\bf The origin of the temperature anisotropies:} what ``seeds'' the temperature inhomogeneities?.
   Simple
  statistical fluctuations are far too small. This can be seen by considering  the simple case of
   a lump of matter with about one solar mass which  has about
  $10^{57}$ particles. Statistical fluctuations will be of
  order $\sim 10^{-23}$ which cannot be reconciled with a
  $\Delta T/T \sim 10^{-5}$. Obviously for much larger masses the
  fluctuations are much smaller.
\end{itemize}

Inflation provides a natural answer to {\em both} questions: the
temperature anisotropies are caused by small inhomogeneities in the
matter density which originated in the quantum fluctuations during the
inflationary era as described in section 3.1.1 above. These density
perturbations are given by eqn. (\ref{delrho}) with physical
wavelengths that are deep inside the Hubble radius and therefore were
in causal contact during inflation (see fig.(\ref{fig:inflation})).
When these wavelengths cross the horizon (becoming super-horizon)
during inflation, they are no longer affected by microphysical
processes, but carry the initial correlations that they had while
inside the Hubble radius. When they re-enter the Hubble radius right
before the LSS, these density fluctuations provide anisotropies in the
gravitational potential. As the photons fall in these anisotropic
potential wells their wavelengths are redshifted, regions with larger
densities provide larger gravitational potential and photons are
redshifted more and are therefore {\em cooler}. These regions were in
causal contact {\em during} inflation, became causally disconnected
from any physical process until they re-entered again the Hubble
radius, carrying the original correlations. Although there are many
details in the relation between temperature and density
anisotropies~\cite{bookliddle,bookcoles,bookkolb} this simple argument
illustrates some of the most important physics for the anisotropies on
the scales measured by COBE. A detailed
analysis~\cite{bookliddle,bookcoles,bookkolb} yields

\begin{equation}\label{deltaT}
\frac{\Delta T}{T_0} \sim \frac{1}{3} \frac{\Delta
\rho}{\rho}\left|_{LSS} \right.
\end{equation}

The {\em main point} of this discussion and the reason for delving on
this subject, is to highlight that the CMB reveals information on
processes that occurred {\em very early}, for example during inflation,
when the physical wavelengths of the perturbations that generate the
temperature anisotropy first crossed the horizon, or {\em very late},
near recombination right after they re-entered. To precisely highlight
this point we have purposely studied in section 3.1 the influence of a
phase transition on the power spectrum of density perturbations, and
therefore by eqn. (\ref{deltaT}) on  the temperature fluctuations. That
is to say, and we emphasize, that if the CMB reveals {\em any}
observable consequences of phase transitions--the main theme of this
lecture-- these are {\em not} part of the standard model: either these
phase transitions ocurred at the time of inflation (perhaps some GUT?)
or at a very low energy scale of ${\cal O}({\rm eV})$. The acoustic
peaks in the multipole expansion of the temperature anisotropies can
give {\em indirect} information on possible phase transitions through
constraints on $\Omega_bh^2$ and the cosmological
parameters~\cite{turner}.

\subsection{Galaxy formation and on to Stars...}

The temperature anisotropies measured in the CMB give the clue to the
origin of large scale structure and the formation of galaxies. As
described above, the temperature fluctuations of the CMB reflect
fluctuations in the density of matter. The currently accepted scenario
for the formation of galaxies and large scale structure begins with the
growth of small (linear) perturbations in the density under the
relentless action of gravity. To understand the main ideas in a
simplified manner, consider a small fluctuation $\delta \rho$ in a
background of constant density $\rho_0$. A region of larger density has
a larger gravitational potential which in turn causes this region to
become denser, i.e, gravity tends to make this overdense region even
more overdense. The dynamical time scale for the free-fall collapse of
this region is

\begin{equation}\label{freefall}
t_{ff} \sim \frac{1}{\sqrt{G_N \rho_0}} \sim \frac{1}{H}
\end{equation}
\noindent with $G_N=1/M^2_{Pl}$ being Newton's gravitational constant.
However a change in density induces a proportional change in pressure
which tends to restore hydrostatic equilibrium. The proportionality
constant is determined by the speed of sound,

\begin{equation}\label{delpress}
\delta p = c^2_s \delta \rho
\end{equation}

Consider a perturbation of wavelength $\lambda$, if the time it takes a
pressure wave to restore hydrostatic equilibrium over this wavelength
is {\em smaller} than the free-fall time (\ref{freefall}), then the
restoring force from the pressure wave prevents the gravitational
collapse. On the other hand if the free-fall time for collapse is {\em
shorter} than the time scale for the pressure wave to restore
equilibrium over the distance $\lambda$, gravitational collapse is
unhindered. The balance between these two time scales determines the
Jeans wavelength

\begin{equation}\label{jeans}
\lambda_J \sim \frac{c_s}{\sqrt{G_N\rho_0}}~~;~~
\left\{\begin{array}{cc}
  \lambda>\lambda_J & \Rightarrow ~{\rm collapse} \\
   \lambda<\lambda_J & \Rightarrow ~{\rm no~ collapse}
\end{array} \right.
\end{equation}
this is the {\em Jeans instability}. In a non-expanding geometry this
instability leads to the exponential growth of density perturbations
for $\lambda > \lambda_J$, but in an expanding geometry the instability
has to catch up with the expansion and as a result small perturbations
grow as a power law with time~\cite{longair}. In particular for a
matter dominated Universe (as is the case after recombination) the
density profile for $\lambda > \lambda_J$ grows in time as
\begin{equation}\label{growthmatt}
\frac{\delta \rho(t)}{\rho_0} \sim a(t) \propto t^{\frac{2}{3}}
\end{equation}
or equivalently
\begin{equation}\label{growthmattII}
\frac{\delta \rho(t)}{\rho_0} \sim (1+z)^{-1}
\end{equation}

To get an idea of the order of magnitude of the Jeans scale it is
illuminating to compute the Jeans mass (mass contained in a volume of
radius $\lambda_J/2$) at the time of recombination. At this time most
of the matter is in the form of hydrogen with an (adiabatic) speed of
sound $c^2_s = 5k_BT/3m_H$ leading to~\cite{longair}

\begin{equation}\label{jeansmass}
M_J = \frac{\pi \lambda^3_J \rho_0}{6} \sim 1.6 \times 10^5
\frac{M_{\odot}}{(\Omega_0h^2)^{\frac{1}{2}}}
\end{equation}
\noindent which is the size of a typical globular cluster.

Since at the LSS $\delta \rho/\rho_0 = 3  \Delta T/T \sim 10^{-4}$ at a
redshift $z_{LSS} \sim 1100$, the argument above suggests that density
perturbations become non-linear (their amplitude becomes of ${\cal
O}(1)$) at redshift of ${\cal O}(1)$, which is consistent with
observations of star-forming regions~\cite{longair}. Thus the
temperature fluctuations in the CMB give direct confirmation of the
paradigm of structure formation based on the gravitational instability
of primordial density perturbations.

The main points of the analysis provided in this section are: i) to
argue that the primordial density fluctuations that determine the
temperature fluctuations of the CMB through (\ref{deltaT}) provide the
seeds for structure formation. Any information from a primordial phase
transition that is imprinted in the anisotropies of the CMB is also
imprinted in the initial power spectrum of density perturbations that
seed large scale structure formation.  ii) to highlight that when the
pressure vanishes, gravitational collapse on large scales goes
unhindered.  This last observation relates to the previous discussion
of the formation of primordial black holes during the QCD phase
transition, which if first order implies an anomalously small (if not
vanishing) speed of sound.

\subsection{Stellar Evolution 101: QGP  at the core of Pulsars?}

The basic Jeans gravitational instability is also at the heart of star
formation. {\em Protostars}  condense from giant molecular (hydrogen)
gas  clouds~\cite{stellar}. A solar mass protostar condenses  in
typical star forming regions with temperatures $T\sim 20 ~K$ and
densities $\gtrsim 10^{-16}~g/cm^3$ the gravitational contraction of
the cloud heats up the gas until the temperature reaches $\sim 10^7~K$
when thermonuclear fusion of hydrogen into helium generates enough
energy and pressure to halt the collapse. The threshold mass for a star
to undergo thermonuclear fusion is about $0.08~M_{\odot}$. Stars with
smaller masses do not attain the temperature to ignite hydrogen and the
gravitational pull is counterbalanced by electron degeneracy, these
``failed'' stars are called {\em brown dwarfs}. Solar mass type stars
live on the main sequence for about 10 billion years and spend most of
their lives burning hydrogen into helium mainly through the three
branches of the pp (proton-proton) chain and more massive stars also
produce helium through the C(arbon)N(itrogen)O(xygen)
chains~\cite{stellar}. When hydrogen is depleted, gravitational
contraction begins again until the temperature reaches about $10^8 K$
when the thermonuclear fusion of helium begins at the core while a
shell of hydrogen still burns outside the core. This helium burning
stage produces carbon (through the miracle of the triple alpha process)
and oxygen. Computer simulations~\cite{stellar} show that stars with
masses up to $\sim 8M_{\odot}$ burn hydrogen and helium but the
temperatures attained at the core are not high enough to ignite carbon.
These stars (the sun is one of them) end up their life cycles as {\em
white dwarfs}, with a degenerate CO core, the gravitational pull is
counterbalanced by the Fermi degeneracy pressure of electrons. Typical
radii and densities of these stars are $R \sim 10^4~{\rm Km};\rho \sim
10^6 gr/cm^3$. The maximum mass for a stellar object to be in
hydrostatic equilibrium balanced by Fermi degeneracy pressure is given
by the Chandrasekhar limit $\sim 1.4~M_{\odot}$. When the density is
such that the Chandrasekhar limit is reached, electrons at the top of
their Fermi surface   become ultrarelativistic and the equation of
state is softer. Stars with masses $> 8-10 ~M_{\odot}$ evolve through
all the stages of nuclear burning beginning with hydrogen, continuing
with helium, carbon, neon, oxygen and silicon burning
(photodisintegration at $T \sim 3.5\times 10^9K$) which ends in the
iron group elements with an ${}^{56}Fe$ core and concentric shells
(``onion structure'') of silicon, neon, oxygen, carbon, helium and
hydrogen. Since Fe has the largest binding energy per baryon,
thermonuclear fusion cannot proceed further. For stars with masses $>
10M_{\odot}$ the iron core reaches the Chandrasekhar limit for support
by electron degeneracy and is on the limit of gravitational collapse.
There are two main factors that trigger the
collapse~\cite{stellar,arnett,synt}: i) the temperature of the core is
now $\sim 8\times 10^9K$ enough to photodisintegrate iron group
elements by the reaction

\begin{equation}\label{photodis}
\gamma+{}^{56}Fe \Leftrightarrow 13 {}^4He+ 4n
\end{equation}

ii) the Fermi energy of the degenerate electrons becomes larger
than the neutron-proton mass difference $\sim 1.3 {\rm Mev}$ and
is therefore large enough for electron capture by protons

\begin{equation}\label{elecap}
e^-+p \rightarrow n+\nu_e
\end{equation}
this is the process of neutronization. As the electrons that support
the gravitational pull with their Fermi pressure are captured,
hydrostatic equilibrium falters and the core begins to collapse on a
dynamical (free fall) time scale given by eqn. (\ref{freefall}). With
the density of the core $\sim 10^9 g/cm^3$ the free fall time scale for
collapse is $\sim 1 {\rm msec}$. As the collapsed core reaches nuclear
matter density, it becomes highly incompressible and its equation of
state stiffens. The infalling matter from the outside shells bounces
from the incompressible core and a shock wave is formed. This shock
blows off the outer layers resulting in  a Supernova (core-collapse or
Type II) explosion. The details of the explosion and the shock are very
complex and involve neutrino transport, convection etc. For more
details the reader is referred to~\cite{stellar,arnett,synt}.

What happens after this depends on the mass of the progenitors.
Current understanding based on numerical
evolution~\cite{stellar,arnett,synt} suggests that for stars with
masses up to about $30M_{\odot}$ the final result is a neutron
star, while for masses larger than this the collapse probably
leads to the formation of a black hole. When the supernova
explosion leads to a neutron star, an important feature with
observational consequences is that during the process of
neutronization of the iron core with the Chandrasekhar mass $\sim
1.4M_{\odot}$ most of the $10^{57} $ protons present in the core
are converted to neutrons and $\sim 10^{57}$ neutrinos each of
about $10 {\rm Mev}$ are released and carry most (up to $99\%$) of
the energy $\sim 10^{53}{\rm ergs}$.  When the density of the
collapsing core reaches about $ 4\times 10^{11}~g/cm^3$, neutrons
begin to {\em drip} out of the nuclei, this is the {\em neutron
drip line} and the collapsing core becomes a gigantic neutron.
Typical neutron stars have masses in the range $1.4-2 M_{\odot}$
with typical radii $\sim 10~{km}$. The most direct evidence for
neutron stars are {\em pulsars} which are believed to be highly
magnetized ($B \sim 10^{12}G$)  rotating neutron
stars~\cite{glen,weber}. Pulsar periods range from about a
millisecond up to a second.

The equation of state for neutron stars for densities {\em below}
nuclear matter $\rho_n \sim 2\times 10^{14}g/cm^3$ is fairly well
understood\cite{heil,prak}, while for densities above $\rho_n$
there is the possibility of hyperon rich matter, pion and or kaon
condensates, muons, and other exotica~\cite{glen,weber,heil,prak}.
Modern theories of superdense nuclear
matter~\cite{heil,prak,glen,weber}, predict a composition for
neutron stars that is depicted qualitatively in
fig.(\ref{fig:nscompo}) below.


\begin{figure}[h]
\begin{center}
\epsfig{file=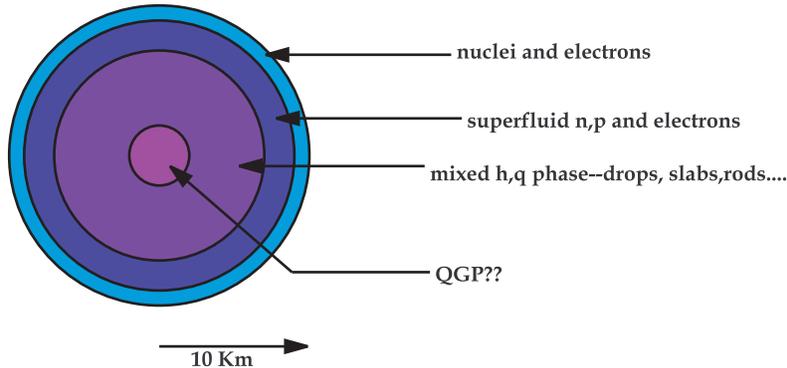}
\caption{Neutron star composition. } \label{fig:nscompo}
\end{center}
\end{figure}


A crust of (${}^{56}Fe$) nuclei, electrons and neutrons, followed by an
internal region of neutron and proton superfluids with the possibility
of pion and kaon condensates and hyperons. But more importantly for
this discussion, current ideas suggest the possibility of a mixed phase
of hadrons and quarks, characterized by geometric structures (a
consequence of the competition between nuclear and Coulomb forces) and
ultimately a  core of {\em deconfined} quarks and gluons with $T \leq
1~{\rm Mev}$ and $\rho \sim 3-5 \rho_n$, i.e, the core is conjectured
to be a {\em cold, dense Quark Gluon Plasma}.

Recently the interesting possibility of novel color superconducting
phases of cold and dense QCD has been proposed and we refer the reader
to several review articles on the subject for details on these
fascinating aspects~\cite{raja}.

In a very tantalizing recent article Glendenning and
Weber~\cite{glenweber} proposed that a phase transition to quark matter
can provide a potential explanation for the anomalous distribution of
pulsar frequencies in low mass X-ray binaries (LMXBs). Observations of
the pulsar frequencies in LMXBs by the Rossi X-ray Timing Explorer
reveal a spike at a frequency $\sim 300~{\rm Hz}$.  We will describe
this fascinating possibility below along with the current experimental
efforts to map the QCD phase diagram with ultrarelativistic heavy ion
collisions.

 The physics is different for Type Ia
Supernovae that are currently at the center of the discussion of their
use as standard candles to study the cosmological
constant~\cite{super1a}. These are conjectured to be CO white dwarfs in
a binary system. The white dwarf accretes mass from its companion until
its own mass becomes just below (a few percent) of the Chandresekhar
limit. Current understanding suggests that at this point the CO core
begins thermonuclear burning under degeneracy conditions. This results
in runaway burning since in the Fermi degenerate situation the pressure
is almost independent of the temperature (just as in the helium flash).
The temperature increase at almost constant pressure leads to a runaway
thermonuclear burning that results in the explosion of the
star~\cite{synt}. What makes these supernovae so special is that their
composition and mass at the time of the explosion is the same for all
Type Ia Supernovae and this feature may explain their similar light
curves~\cite{brian}.

\subsection{Executive Summary: observational consequences of cosmological phase transitions?}

The focus of the review of big bang cosmology as well as the
astrophysics of (compact) stars is to highlight where and when there
could be observational consequences of the phase transitions predicted
by the standard model of particle physics. The picture that emerges
from this brief tour through the early Universe, is that while current
CMB measurements and large scale surveys have the potential for
revealing evidence for {\em inflationary} phase transitions the
observable evidence for a standard model phase transition is at best
{\em indirect}. I have argued that inflationary phase transitions,
perhaps taking place at the GUT scale could leave an imprint in the
power spectrum of temperature anisotropies, for example a red tilt as a
consequence of spinodal instabilities~\cite{inflationPT}. Or as argued
in ref.~\cite{sarkar} a step-like feature in the power spectrum of
temperature anisotropies or in the power spectrum of galaxy clusters
may be produced by an inflationary phase transition in a supersymmetric
theory. Either proposal however, lies beyond the {\em standard model}
and at energy scales unlikely to be probed by any current or future
accelerator. The main ``problem'' with detecting phase transitions with
the CMB or large scale surveys is that the fluctuations that can seed
 temperature and density inhomogeneities were generated very {\em
early}, during inflation, i.e, at an energy scale $\sim
10^{16}~{\rm Gev}$, or very {\em late} during recombination, i.e,
at an energy scale $\sim 0.3~{\rm eV}$. As is clear from
fig.(\ref{fig:inflation}) scales of cosmological relevance today,
say from $\sim 10$Mpc up to the Hubble radius $\sim 3000$Mpc, were
super-horizon, hence decoupled from the microphysics at the time
of the EW or QCD phase transitions.

The {\em indirect} evidence for standard model phase transitions
is contained in the baryon to photon ratio, i.e, the baryon
asymmetry and in the possibility of primordial black holes or
inhomogeneous nucleosynthesis caused by the QCD phase transition.
As argued above, it now seems clear that the standard Electroweak
theory cannot describe consistently baryogenesis because of the
smallness of the CP violating parameter(s) as well as the large
Higgs mass. This analysis leaves the responsibility for
observational consequences of standard model phase transitions
squarely on the shoulders of QCD!. A strong first order phase
transition from a QGP to a hadron phase can lead to the formation
of  primordial black holes because the speed of sound becomes
anomalously small thus gravitational collapse of solar mass clumps
is unhindered. There is the possibility of inhomogeneities in the
baryon density leading to inhomogeneous nucleosynthesis with the
tantalizing possibility of explaining the new bounds on $\Omega_b
h^2$ from Boomerang and Maxima. There is also the possibility for
the formation of strangelets or strange quark nuggets that could
be a component of the cold dark matter. As described above most of
these possibilities rely on the details of the QCD PT and the
parameters of QCD that determine the main dynamical aspects of the
transition. While the lattice gauge theories program has the
potential for computing many of these quantities, it is clearly
important to study {\em experimentally} the phase transitions of
the standard model. As mentioned above the {\em direct study} of
the EWPT is certainly out of the reach of current and most
certainly near future accelerators: to achieve the PT temperature
$T \sim 100~{\rm Gev}$ requires an energy density $\varepsilon
\sim 10^{11}\rho_n$. The QCD phase transition is perhaps the {\em
only} phase transition of the standard model that {\em can and is
being} studied with accelerators. At the QCD PT temperature $T
\sim 200 ~{\rm Mev}$ the energy density required is $\varepsilon
\sim 2-4 \rho_n$ which is the energy density achieved in {\em
ultrarelativistic heavy ion collisions}. I now describe the
current (and future) program to study the QCD phase transition(s)
with ultrarelativistic heavy ion collisions, as well
 the recent fascinating suggestion that there could be
observable hints of quark matter in the core of pulsars.

\section{Relativistic heavy ion collisions and pulsars open a window  to the early
Universe:}

The program of relativistic heavy ion collisions whose primary goal is
to study the phase diagram of QCD began almost two decades ago with the
fixed target heavy ion programs at the AGS at Brookhaven and the SPS at
CERN. Recently~\cite{heinz,braun} a summary of the results  of these
efforts mainly through the $Pb+Pb$ experiments at SPS-CERN provided
very exciting ``evidence''  in favor of the existence of the QGP. As
impressive as this  body of evidence is, the consensus in the field is
that while the evidence is very suggestive it is far from conclusive.
As SPS  shuts down to pave the way for the forthcoming Large Hadron
Collider wherein the Alice program will continue the search for the
QGP,  the torch now passes to the R(elativistic) H(eavy)I(on)C(ollider)
at BNL.

\subsection{RHIC and LHC seek the QGP: the big picture}
The Relativistic Heavy Ion Collider (RHIC) at BNL is currently studying
$Au+Au$ collisions with center of mass energy $\sqrt{s} \sim 200 {\rm
AGev}$ and luminosity $\sim 10^{26} cm^{-2}s^{-1}$. The future ALICE
heavy ion program at LHC is expected to study $Pb+Pb$ collisions with
c.m. energies up to  $\sqrt{s}\sim 5 {\rm ATev}$ and luminosities $\sim
10^{27} cm^{-2}s^{-1}$. In these collisions the heavy nuclei can be
pictured in the CoM frame as two Lorentz contracted pancakes, for
example for $Au+Au$ collisions the size of each ``pancake'' in the
direction transverse to the beam axis is about 7 fm. At RHIC and LHC
energies most of the baryons are expected to be carried away by the
receding pancakes (the fragmentation region) while in the region of the
collision a large energy (density) is deposited in the form of quark
pairs and gluons. At least two important mechanisms for energy
deposition in the collision region are at
work~\cite{muharris,blaizot,qgpbooks}: i) the establishment of a strong
color electric field (flux tube) that eventually breaks up into
quark-antiquark pairs when the energy in the field is larger than the
pair production threshold and ii) the partons (quarks and gluons)
inside the colliding nuclei interact and redistribute their energy in a
``parton cascade''~\cite{geiger}. An estimate of the energy deposited
in the collision region has been provided by Bjorken~\cite{bjorken}

\begin{equation}\label{enercen}
\epsilon = \frac{1}{\tau_0 \pi R^2_A} \frac{dE_T}{dy}
\end{equation}

\noindent with $\tau_0 \sim 1 {\rm fm}/c$, $R_A \sim 7 {\rm fm}$ for Au
and $dE_T/dy$ is the transverse energy per unit rapidity which is {\em
measured}. At RHIC the transverse energy per unit rapidity is expected
to be in the range $500-900 {\rm Gev}$, giving for the energy density
$\epsilon \sim 4-6 {\rm Gev}/{\rm fm}^3$ translating this value into
the temperature of a quark-gluon gas leads to $T\gtrsim 400 {\rm Mev}$
larger than the expected critical temperature $T_c \sim 160 {\rm Mev}$.
The evolution of the parton distribution functions reveals that
thermalization of quarks and gluons occurs on time scales ${\cal
O}(0.3-0.7) {\rm fm}/c$ with gluons thermalizing
first~\cite{muharris,blaizot}. After the quark-gluon plasma achieved
LTE, the evolution is conjectured to be described by hydrodynamic
expansion~\cite{hydro}. As the QGP expands and cools, the temperature
falls near the critical temperature and the confinement and chiral
phase transitions occur. Upon further cooling the quark-gluon plasma
hadronizes, the hadrons rescatter until the hadron gas is dilute enough
that the mean free path is larger than the mean distance between
hadrons. At this point hadrons ``freeze-out'' and stream out freely to
the detectors from this {\em last scattering or freeze-out surface}.
This picture is summarized in fig. (\ref{fig:uhic}) below.


\begin{figure}[h]
\begin{center}
\epsfig{file=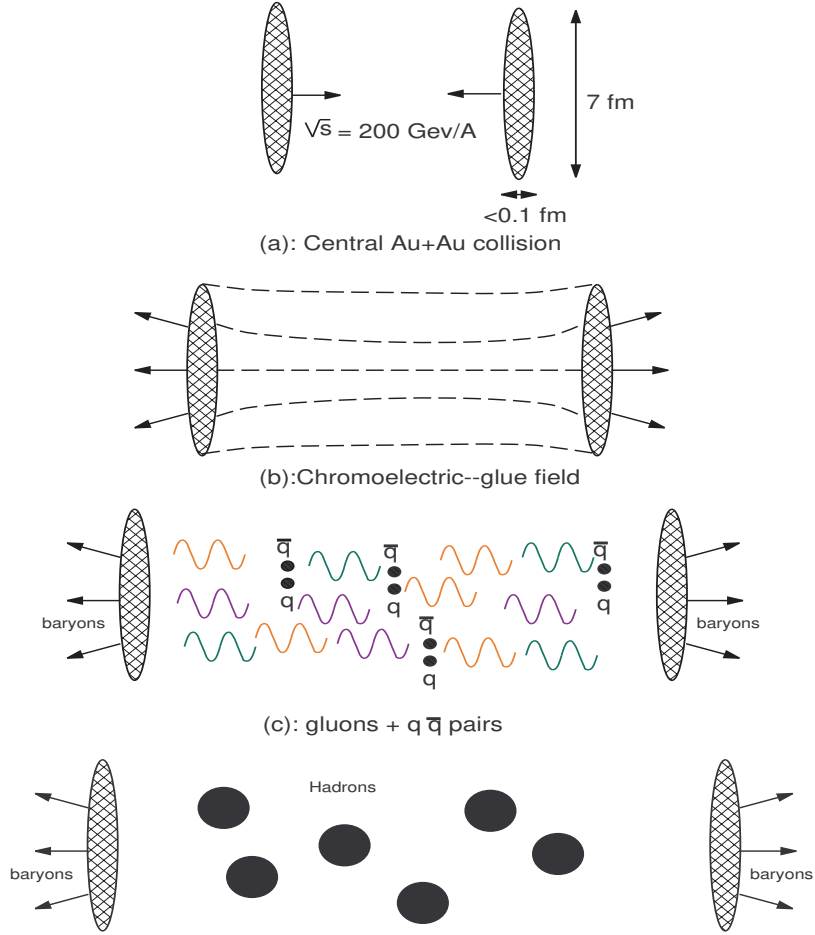,height=5in,width=5in}
\caption{Heavy Ion Collisions. } \label{fig:uhic}
\end{center}
\end{figure}


Current estimates based on this picture and on detailed numerical
evolution~\cite{evolutionqgp} suggest that at RHIC the quark gluon
plasma lifetime is of order $\sim 10 fm/c$ while the total evolution
until freeze-out is $\sim 50-100 fm/c$.

\subsubsection{Hydro, LGT and the EoS:}
Although the evolution of the quark gluon plasma from the initial
state described by the parton distribution of the colliding nuclei
until freeze-out of hadrons  clearly requires a non-equilibrium
description, a hydrodynamic picture of the evolution is both
useful and experimentally relevant~\cite{hydro}. In the
approximation in which quarks and gluons are strongly coupled in
the sense that their mean free paths are much smaller than the
typical wavelength for the variation of collective phenomena, the
QGP can be described as a {\em fluid} in LTE. Hydrodynamics
results from the conservation laws applied to the fluid form of
the energy momentum tensor and the conserved currents. In
particular for a fluid in LTE, the energy momentum tensor is of
the form
\begin{equation}\label{tmunu}
T^{\mu \nu} = (\varepsilon+p)u^{\mu}u^{\nu}-pg^{\mu \nu}
\end{equation}
\noindent with $\varepsilon,p$ the energy density and pressure
respectively, $g^{\mu \nu}$ is the  metric and

\begin{equation}\label{velo}
u^{\mu}=\gamma(x)(1,{\vec v}(x))
\end{equation}

\noindent is a local 4-velocity vector with $\gamma(x)$ the local
Lorentz contraction factor. In this description the dynamical evolution
of the fluid is obtained from the conservation laws of energy-momentum,
baryon number and entropy~\cite{qgpbooks,hydro} and to close the set of
equations one needs an equation of state $p=p(\varepsilon)$.

In order to make a definite connection with the cosmological setting as
well as for it simplicity and usefulness to describe qualitatively the
expansion and cooling of the QGP we now briefly summarize Bjorken's
model for longitudinal expansion~\cite{qgpbooks,bjorken,hydro}. There
are three important ingredients in Bjorken's models: i) the ``central
rapidity'' region where the energy deposited and the formation of the
QGP takes place is well separated from the ``fragmentation region''
which is the region of the receding pancakes where most of the baryons
are (see fig.(\ref{fig:uhic}). ii) The thermodynamic variables are
invariant under boosts along the beam (longitudinal) axis, this is
based on the observation that the particle distributions are invariant
under these boosts in the ``central rapidity region'' and  iii) the
expansion only occurs along the beam axis, i.e, longitudinal expansion.
In the baryon free region, local thermodynamics implies that the
entropy density in the local rest frame of the fluid is related to the
energy density and pressure as
\begin{equation}\label{entropy}
\varepsilon+p = Ts
\end{equation}

It is convenient to introduce the space time rapidity variable $\eta$
and proper-time $\tau$

\begin{eqnarray}\label{etatau}
\eta =  \frac{1}{2}\ln\left[\frac{t+z}{t-z} \right] ~~;~~\tau
=\sqrt{t^2-z^2}
\end{eqnarray}

\noindent with $z$ the coordinate along the beam axis (longitudinal).
Introducing the fluid rapidity $\theta$ by writing the local velocity
(\ref{velo}) as

\begin{equation}\label{fluidrap}
u^{\mu} = (\cosh(\theta),0,0,\sinh(\theta))
\end{equation}

Furthermore Bjorken's model assumes the fluid to be composed of free
streaming particles for which $v_z=z/t$ in which case the fluid
rapidity $\theta$ becomes the space-time rapidity $\eta$. Boost
invariance along the longitudinal direction entails that
$\varepsilon,p,s,T$ are all functions of proper time only (here s,T)
are the entropy density and temperature~\cite{bjorken,hydro}.

The energy-momentum conservation equation $\partial_{\mu}T^{\mu \nu}$
leads to two equations by projecting along the direction $u^{\mu}$ and
perpendicular to it using the projector $g^{\mu\nu}-u^{\mu}u^{\nu}$.
Under the assumption that the central region is baryon free, the
conservation of entropy and energy and momentum lead to the following
equations (for details see~\cite{hydro,meyer,qgpbooks,bjorken})

\begin{eqnarray}
&&\frac{d\varepsilon(\tau)}{d\tau}+\frac{1}{\tau}(\varepsilon+p)=0 \label{enercons}\\
&&\frac{ds(\tau)}{d\tau}+\frac{s}{\tau}=0 \label{entrocons}
\end{eqnarray}

Assuming an equation of state $p(\tau)= p(\varepsilon(\tau))$ and
combining eqns. (\ref{enercons},\ref{entrocons}) with the thermodynamic
relation  (\ref{entropy}) (for baryon free plasmas) one finds that the
evolution of the temperature is given by
\begin{equation}\label{temp}
\frac{\tau}{T}\frac{dT}{d\tau}=-c^2_s ~~; ~~ c^2_s=
\frac{dP}{d\varepsilon}\left|_s \right.
\end{equation}

Which for constant speed of sound results in the cooling law

\begin{equation}\label{coollaw}
T(\tau) = T_0 \left(\frac{\tau_0}{\tau} \right)^{c^2_s}
\end{equation}

Obviously the form of the equation (\ref{enercons}) is similar to the
energy conservation equation in a homogeneous and isotropic metric
given by (\ref{conener}) with the expansion rate $\dot{a}/a =1/3\tau$
when the proper time $\tau$ is identified with the comoving time $t$ in
the cosmological setting. The similarity becomes even more remarkable
for the case of the QGP being modelled as a radiation fluid (which is
expected to be a good approximation at high temperature) since in this
case $c^2_s=1/3$ and the connection with a radiation dominated
cosmology with scale factor $a(t)\propto t^{\frac{1}{3}}$ is evident.

To find the general evolution equations for the QGP under the
conditions specified above, an equation of state is needed.

\begin{figure}[h]
\begin{center}
\epsfig{file=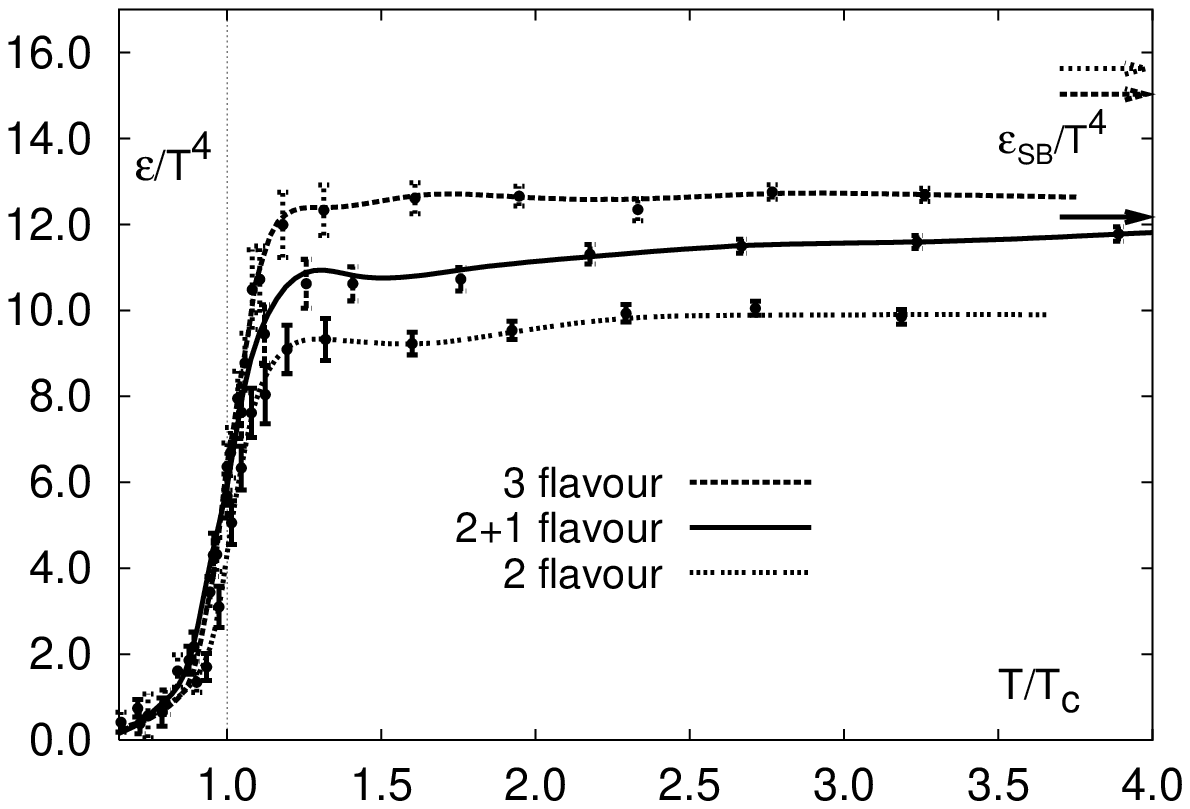,height=2in,width=2in}
\epsfig{file=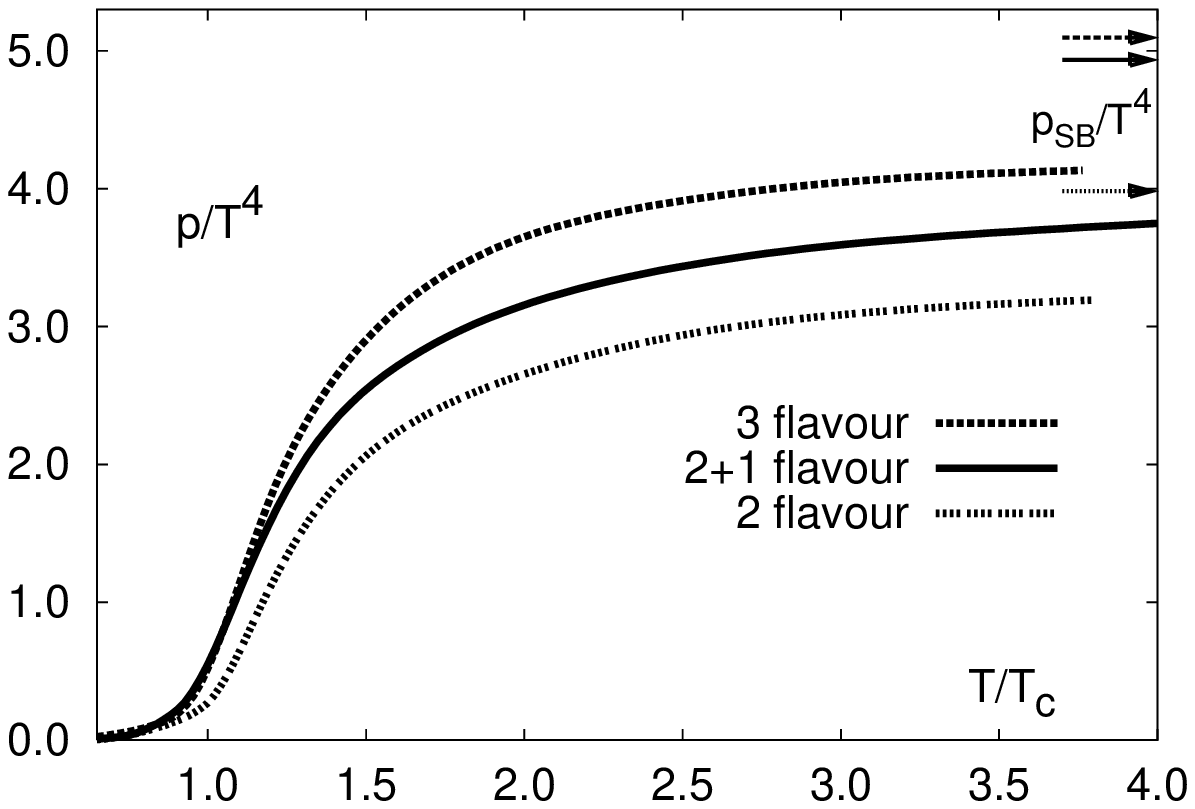,height=2in,width=2in}
 \caption{$\varepsilon/T^4$ and $p/T^4$, vs $T/T_c$ for a $16^3\times
 4$ lattice.From~\cite{karsch2}. } \label{fig:lgt}
\end{center}
\end{figure}

 It is at this stage where the connection with the lattice gauge
theory (LGT) program is made. LGT obtains the thermodynamic
functions {\em in equilibrium} and these are input in the
hydrodynamic description as local functions of space-time under
the assumption of LTE. Fig. (\ref{fig:lgt}) below is an  example
of recent results from LGT for the energy density and pressure
(both divided by $T^4$ to compare to a free gas of massless quarks
and or gluons) as  a function of $T/T_c$ with $T_c=170 ~{\rm
Mev}$.

It is clear from these figures that there is a sharp decrease in the
energy density and pressure at $T=T_c$, the value of the energy density
at $T_c$ is, from the figure $\varepsilon_c \sim 6 T^4_c$. Furthermore
the large temperature behavior is {\em not quite} given by the
Stephan-Boltzmann law ((SB) in the figures), suggesting that even at
large temperatures the plasma is not described by free quarks and
gluons up to temperatures $T \sim 4.0 T_c \sim 700 ~{\rm Mev}$. This
discrepancy is currently explained in terms of quasiparticles rather
than free particles (see below). Nevertheless, what transpires clearly
from these figures is that for temperatures $T >T_c$ the plasma is
approximately a gas of almost free massless quarks and gluons, while
for $T<<T_c$ the plasma is dominated by heavy particles. This is the
picture that deconfinement-confinement phase transition suggests.

A hydrodynamic description combined with the non-perturbative
lattice gauge theory program to extract the EoS of the QGP provide
a basic quantitatively reliable picture of the dynamical evolution
of the QGP. The approximation of ideal (or perfect) hydrodynamics
can be relaxed by considering {\em viscous} hydrodynamics by
including the bulk and shear viscosity in the energy momentum
tensor~\cite{hydro,gyul}.

\subsection{Thermalization, quasiparticles and the EoS:}

The initial stages of evolution of the QGP beginning from the
distribution of partons (quarks and gluons) in the colliding nuclei is
studied with a semiclassical transport approach that includes
perturbative QCD cross sections with screening
corrections~\cite{geiger}. These studies reveal that gluons and quarks
thermalize on time scales $< 1 fm/c$ with gluons thermalizing first.
While a classical (or semiclassical) transport approach may not be
completely justified in this regime, a qualitative and quantitative
idea of the relaxation time scales can be obtained by {\em assuming} a
thermalized state and studying how a small departure from equilibrium
relaxes. The relaxation rate is given by $\Gamma = n\sigma$ with $n$
the particle density and $\sigma$ a typical scattering cross section.
Consider the relaxation of quarks in a bath of gluons at temperature
$T$, one gluon exchange yields a typical cross section

\begin{equation}\label{crosssec}
\sigma \sim \frac{\alpha^2_s}{T^2}
\end{equation}
\noindent with $\alpha_s = g^2/4\pi$ and $g$ is the gauge coupling
constant and we have assumed that the typical energy of the exchanged
gluon is $\sim T$. For $T
>>m$ with $m$ the typical masses, $n\propto T^3$ leading to

\begin{equation}\label{gammastrong}
\Gamma \sim \alpha^2_s T
\end{equation}

Gluons of typical energy $T$ are hard and lead to smaller cross
sections, while the exchange of soft gluons leads to larger cross
sections and  is more efficient for thermalization. However soft
gluons are screened by the medium.  The estimate above, eqn.
(\ref{crosssec}) is reliable for hard gluon exchange, but
overlooks the fact that {\em soft} longitudinal gluons are Debye
screened. A careful calculation~\cite{lebellac} reveals that the
Debye screening length $m_D \sim gT$ therefore the denominator in
eqn. (\ref{crosssec}) must be replaced by $T^2 \rightarrow
g^2T^2$. Furthermore, while longitudinal (electric) gluons are
Debye screened, transverse (magnetic) gluons are only {\em
dynamically screened} by Landau damping~\cite{brapis,lebellac}.
Including these screening effects it is found that the quark
relaxation rate is given by~\cite{lograte}

\begin{equation}\label{quarkrel}
\Gamma \sim \alpha_s T \ln\left(\frac{1}{g}\right)
\end{equation}

\noindent for $\alpha_s \sim 0.3; T \sim 300 {\rm Mev}$ one finds the
typical relaxation time $\tau_{rel}= \Gamma^{-1} \sim 0.8-1 fm/c$.

The main point in bringing up this estimate is to highlight the
fact that in a medium there are many-body effects that screen or
``dress'' the particles, and as a result, even for high
temperatures the plasma is {\em not} described by a free gas of
quarks and gluons, but in terms of ``dressed'' weakly interacting
{\em quasiparticles}. In a thermalized QGP these quasiparticles
have a typical ``thermal mass'' of order $gT$~\cite{lebellac}. A
framework to include these screening corrections consistently in
the perturbative expansion has been put forth in
ref.~\cite{brapis} and is referred to as the hard thermal loop
(HTL) program. In fact the energy density and pressure obtained by
the LGT program depicted in fig. (\ref{fig:lgt})  have been
recently reproduced analytically in terms of {\em thermal
quasiparticles} with a typical mass of order $gT$~\cite{quasi}.
Thus the combination of numerical results from LGT and the
analytic resummation provided by the HTL program lead to a picture
of the QGP at high temperatures in terms of weakly interacting
quasiparticles with thermal masses $\sim gT$.

\subsection{Predictions and observations:}

Several experimental signatures had been associated with the formation
of the QGP~\cite{muharris,signatures}, and the heavy ion programs had
focused on several of them. Here I summarize the observables that have
been proposed as telltales of a QGP and the data gathered by the SPS
from $Pb+Pb$  and $Pb+Au$ collisions, which taken together provide a
hint of evidence~\cite{heinz,braun,blaizot} for the effects associated
with a QGP (although many if not all of them could have alternative
explanations).

\subsubsection{$J/\Psi$ suppression:}
The $J/\Psi$ is a $\bar{c}c$ bound state that is very narrow and that
if produced in the early stages of the collision can probe the QGP
because its lifetime is longer than that of the QGP and decays into
dilepton pairs which leave the plasma without scattering. The original
suggestion~\cite{satz} is that when the screening length of the color
force is smaller than the size of the $\bar{c}c$ bound state, this
narrow resonance will ``melt''. This original argument thus suggests
that a suppression of the $J/\Psi$ yield could provide information on
the existence of the QGP. Alternatively a suppression could result from
scattering with hard gluons in the plasma and the ``dissociation'' of
the bound state~\cite{satz}. A normal suppression of charmonium is
expected  on the grounds that once formed the $\bar{c}c$ bound state
interact with other nucleons inside the nucleus. This expected
suppression is studied in proton nucleon collisions and extrapolated to
nucleon-nucleon collisions. This is considered the ``normal''
suppression in contrast to the ``abnormal'' suppression expected from
the presence of a plasma. Fig. (\ref{fig:jpsisup}) below shows the data
gathered by the NA50 collaboration at the SPS-CERN~\cite{jpsi}.
\begin{figure}[h]
\begin{center}
\epsfig{file=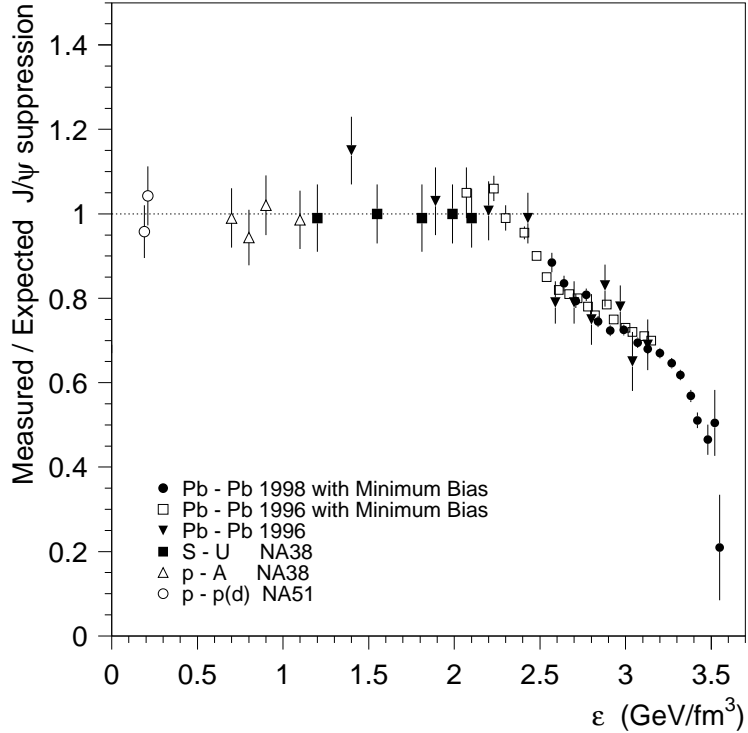,height=4in,width=4in}
\caption{Anomalous $J/\psi$ suppression as a function of the
initial energy density. From~\cite{jpsi}}. \label{fig:jpsisup}
\end{center}
\end{figure}


This striking figure clearly reveals an ``abnormal'' suppression when
the energy density is $\varepsilon > 2.5 {\rm Gev}/{\rm fm}^3$. The
energy density in this figure has been computed with Bjorken's formula
(\ref{enercen}). Further analysis~\cite{heinz,braun} reveals that this
anomalous suppression cannot be accounted for by collisional
dissociation through final state hadronic interactions.

A recent report from the NA50 collaboration at CERN-SPS~\cite{jpsi2}
presents the combined data for $J/\Psi$ suppression from the NA38 and
NA50 experiments. The analysis of the data reveals that while for the
most peripheral (largest impact parameter) collisions the suppression
can be accounted for by nuclear absorption, there is no saturation in
the suppression in the  most central $Pb+Pb$ collisions and that the
observed suppression pattern can be naturally understood in a
deconfinement scenario. This report concludes that {\em the $J/\Psi$
suppression pattern observed in the NA50 data provides significant
evidence for deconfinement of quarks and gluons in $Pb+Pb$ collisions}.

\subsubsection{Electromagnetic probes: dileptons and direct
photons}

Electromagnetic probes: $e^+e^-$  or $\mu^+\mu^-$ dilepton pairs
and direct (prompt) photons are prime probes of the hot
plasma~\cite{electro}, since once produced they leave the plasma
without further interactions because their mean free path is {\em
much} larger than the typical size of the plasma.

{\bf Dileptons:}

The $\rho$ vector meson has particular relevance in this regard
because it decays into dileptons and its lifetime is $\sim 1
fm/c$, therefore once it is produced in the hadron gas it decays
{\em within} the hot hadronic plasma and the produced dileptons
carry directly this information. Thus while dileptons produced
from the decay of the $\rho$ meson do not yield evidence of the
earlier stages in the QGP, they do nevertheless offer information
on the hadronic stage. Figure (\ref{fig:ceres}) below presents the
data gathered by the CERES-NA45 collaboration for the invariant
mass spectrum for electron-positron pairs from 158 AGev $Pb+Au$
collisions at the SPS-CERN.


\begin{figure}[h]
\begin{center}
\epsfig{file=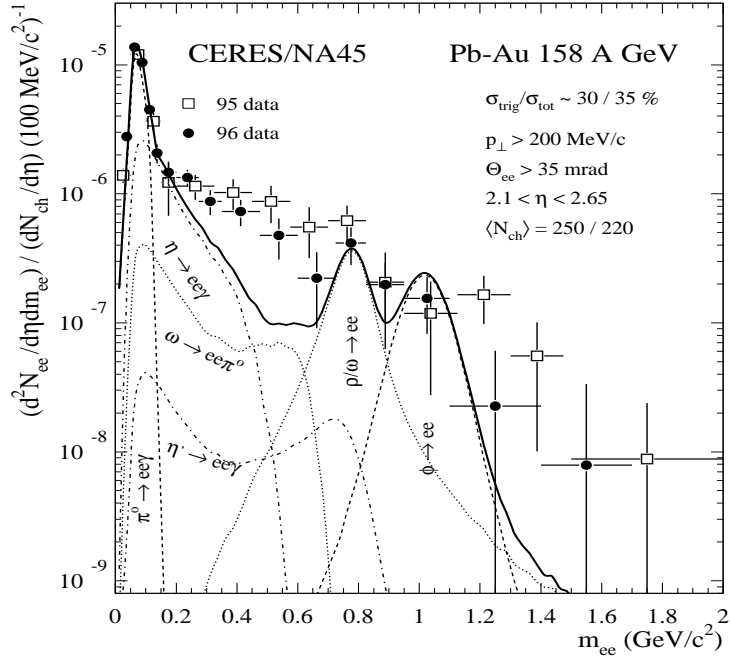,height=4in,width=4in}
\caption{Invariant mass spectrum of $e^+e^-$ pairs from $Pb+Au$
collisions at 158 AGev at the SPS-CERN (CERES/NA45).
From~\cite{ceres}. } \label{fig:ceres}
\end{center}
\end{figure}


The solid line represents the expected spectrum from the decays of
hadrons produced in proton-nucleon  and proton-proton collisions
extrapolated to $Pb+Au$ collisions and is the sum of the contributions
shown in the graph.  There are two  remarkable features in this graph:
a clear enhancement of dileptons in the region $250~{\rm Mev} \lesssim
M_{e^+e^-} \lesssim 700~{\rm Mev}$ and  that instead of the $\rho$
meson peak at $m_{\rho}=770 ~{\rm Mev}$ there is a broad distribution.
The excess of dileptons in the small invariant mass region cannot be
explained by charged pion annihilation~\cite{heinz,braun}. What is
remarkable in this data is that the dilepton enhancement is {\em below}
the putative $\rho$ peak and that there is no hint of the $\rho$ at
$770~{\rm Mev}$!. The current understanding of these features is that
the medium effects result in a shift in the $\rho$ meson mass as well
as a change in its width\cite{wam,heinz,braun,blaizot}. Thus while this
interpretation does not directly yield information on the QGP, it does
support the picture of a hot gas of hadrons, mainly pions which is the
main interaction channel of the $\rho$ vector meson.

{\bf Direct photons: }

Direct photons are conceptually a clean direct probe of the early
stages of the QGP. Photons are produced in the QGP by several
processes: gluon-to-photon Compton scattering off (anti)quark
$q(\bar{q})g\rightarrow q(\bar{q})\gamma$ and quark-antiquark
annihilation to photon and gluon $q\bar{q}\rightarrow g\gamma $ and to
the {\em same order} (see Kapusta et. al. and Aurenche et. al.
in~\cite{electro}) (anti)quark bremsstrahlung $qq(g)\rightarrow
qq(g)\gamma$ and quark-antiquark annihilation with scattering
$q\bar{q}q(g)\rightarrow q(g)\gamma$. Detailed calculations including
screening corrections~\cite{electro} reveal that direct photons from
the QGP could provide a signal that could be discriminated against that
from the hadronic background. This work indicates the theoretical
feasibility of direct photons as direct probes of the early stages of
the QGP.

Recently the WA98 collaboration at SPS-CERN reported their
analysis for the {\em first} observation of direct photons from
$Pb+Pb$ collisions with $\sqrt{s}= 158 AGev$~\cite{WA98}. Their
data is summarized in the fig.(\ref{fig:photons}) below.


\begin{figure}[h]
\begin{center}
\epsfig{file=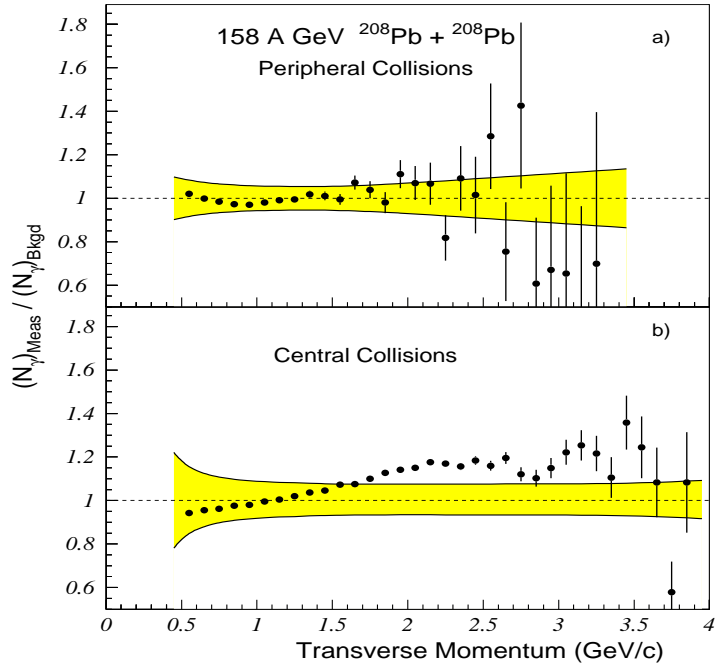,height=4in,width=4in} \caption{Ratio
of total measured yield of direct photons to hadronic background
vs. transverse momentum from $Pb+Pb$ at 158 AGev for peripheral
and central collisions. From~\cite{WA98}. } \label{fig:photons}
\end{center}
\end{figure}


The transverse momentum distribution of direct photons is
determined on a statistical basis and compared to the background
photon yield predicted from a calculation of the radiative decays
of hadrons. The most interesting result is that a significant
excess of direct photons beyond that expected from proton-induced
reaction at the same $\sqrt{s}\,$ is observed in the range of
transverse momentum greater than about $1.5\;{\rm GeV}/c$ in
central collisions.

A detailed analysis of the data and the theoretical expectations based
on the calculations of direct photons from an equilibrated QGP was
recently performed~\cite{hatsuda}. The conclusions of that analysis is
that while it is not clear if SPS has reached the energy density to
form the QGP, the data supports indications of a hot and dense phase
that could be the precursor of the QGP.

More recently~\cite{wangnon} it has been suggested that non-equilibrium
effects in an expanding QGP formed in RHIC collisions could lead to an
enhancement of direct photons in the region of transverse momentum $p_T
\gtrsim 2~ {\rm Gev}$.

\subsubsection{Strangeness enhancement}

Strangeness enhancement along with chemical equilibration are some of
the earliest proposals for clear signatures of the formation of a
QGP~\cite{strangeness}. The main idea is based on the estimate that the
strangeness equilibration time in a hot QGP is of the same order as the
expected lifetime of the QGP ($\sim 10 fm/c$) produced in
nucleus-nucleus collisions. Two important aspects of this estimate make
strangeness enhancement a prime candidate: if strangeness attains
chemical equilibrium in the QGP, this equilibrium value is
significantly higher than the strangeness production in nucleon-nucleon
collisions. Also strangeness production through hadronic rescattering
or final state interactions was estimated to be negligibly
small~\cite{strangeness}. In the QGP, color deconfinement leads to a
large gluon density that leads to the creation of $s\bar{s}$ pairs,
furthermore chiral symmetry makes the strange quark lighter thus
lowering the production threshold. This situation is in contrast to the
case of hadronic rescattering or final state interactions where the
production of pairs of strange quarks has large thresholds and small
cross sections~\cite{heinz,braun}. The usual measure of strangeness
enhancement is through the ratio

\begin{equation}\label{strangeratio}
\lambda_s = \frac{2 \langle \bar{s}s \rangle}{\langle
\bar{u}u+\bar{d}d \rangle}
\end{equation}

Fig. (\ref{fig:strange}) below displays $\lambda_s$ as a function
of $\sqrt{s}$ the energy of the collision for nucleon-nucleon as
well as nucleus-nucleus collisions ($S+S,S+Ag,Pb+Pb$) at SPS.


\begin{figure}[h]
\begin{center}
\epsfig{file=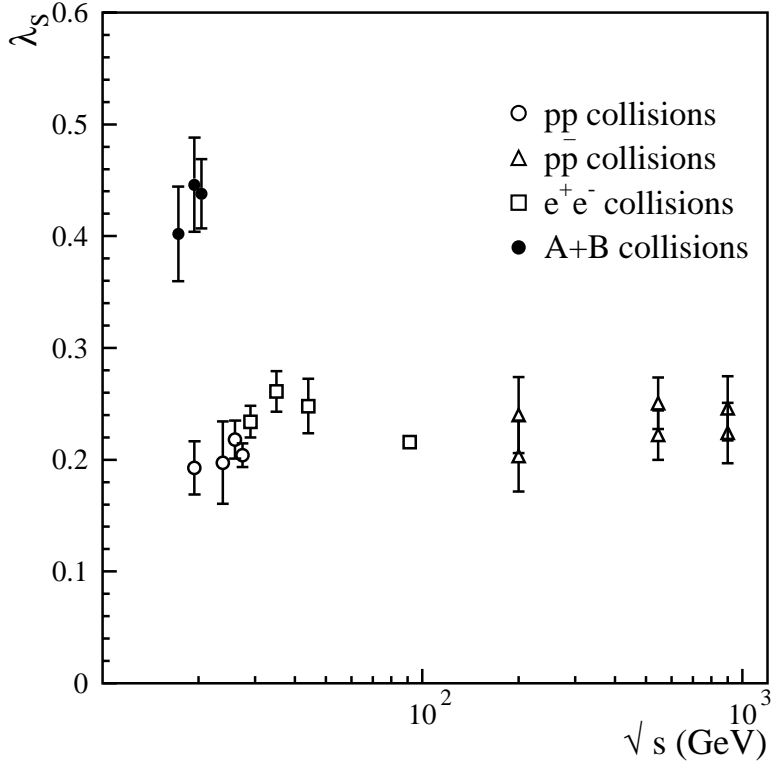,height=4in,width=4in}
\caption{The ratio $\lambda_s$ as a function of $\sqrt{s}$, the
energy of the collision for nucleon-nucleon and nucleus-nucleus
collisions.  From~\cite{soll}} \label{fig:strange}
\end{center}
\end{figure}


It is clear that nucleus-nucleus collision is creating a different
environment that enhances the formation of strangeness. As
mentioned above, and confirmed by detailed numerical evolution of
kinetic models~\cite{kine}, the enhancement {\em cannot} be
explained by hadronic scattering. Thus the enhancement displayed
by fig. (\ref{fig:strange}) provides a very exciting hint that
nucleus-nucleus collisions at SPS are creating a hot state of
matter.

\subsubsection{Collective flow and the Eos:}

 After the collision and the formation of the QGP, the pressure in
 the plasma drives its expansion and cooling. Pressure gradients
 give rise to changes in the velocity distribution and the
 collective motion of the plasma which in turn characterize the
 dynamical  spatial and momentum correlations of the particles in
 the plasma. The collective motion of the plasma driven by the
 internal pressure is referred to as {\em collective flow}. This
 flow provides information on the equation of state (Eos) of the
 plasma as is manifest in the hydrodynamic
 description described in section 4.1.1 above.

 The collective radial expansion of the plasma is typically
 assessed by looking for deviations of momentum distributions from
 thermal although perhaps a more clear assessment is
 obtained by comparing spectra of particles with different masses~\cite{daniel}.

 If the transition from the QGP to the hadron gas phase is first
 order and occurs in LTE, then the two phases coexist at the same
 temperature and pressure  is constant through the transition (through the Maxwell construction).
  However the
 ratio of pressure to energy density decreases and reaches a {\em
 minimum}  at a particular energy density $\varepsilon_{sp} \sim
 1.4 Gev/fm^3$ known as the {\em softest point of the
 Eos}~\cite{shuryak}. As the energy density passes through the
 softest point the effective speed of sound becomes anomalously
 small, and the small pressure cannot accelerate effectively the
 matter and the flow stalls. There is a host of numerical
 simulations based on dynamical transport approaches that predict
 noticeable variations in collective flow for different equations
 of state (see for example~\cite{daniel,shuryak}).

The measurement of collective flow and an assessment of the EoS is
 an integral part of the experimental program at SPS and RHIC.
 There are already interesting results that prove the viability of
 the study of flow to understand the dynamical evolution of the
 hadronic component~\cite{flowex}. But while there is an important
 and exciting
 body of results from SPS and some recent results from STAR at
 RHIC~\cite{flowex} it is still rather difficult and perhaps
 premature to extract clear information on the nuclear EoS.

 The main point to bear, however, is that the EoS is {\em
 experimentally accessible through measurements of flow}. Hence
 the experimental program in relativistic heavy ion collisions has
 access to the EoS which as described above is a very important aspect of the QCD phase
 transition.

\subsubsection{Other predictions...}

There are a variety of other ``predictions'' that purport to describe
potentially observable signatures of a QGP and we mention here a few
that could also be of potential relevance for early Universe cosmology.

As mentioned above, there are {\em two} phase transitions ocurring at
about the same temperature, the confinement-deconfinement and the
chiral symmetry breaking phase transition. Most of the observables
described above refer to the confinement-deconfinement phase
transition. The chiral phase transition refers to the breaking of the
(approximate) symmetry corresponding to independent rotations of the
right and left handed components of the u,d quarks. The three pions are
the (quasi) Goldstone bosons resulting from the breakdown of chiral
symmetry.

 If
the chiral phase transition occurs out of equilibrium, it is
conceivable that strong fluctuations of the pion field could emerge.
These fluctuations had been given the generic name of ``disoriented
chiral condensates'' (or DCC's) ~\cite{dcctheory} and while there could
be different manifestations of these DCC's, all bear in common the
notion of large amplitude, coherent pion ``domains''. These large
fluctuations would result in large regions in which isospin and
probably charge are correlated. Furthermore if these domains are
produced via long-wavelength (spinodal) instabilities, such as those
mentioned within the context of a supercooled phase transition in
section 3.1 above (see the discussion below eqn. (\ref{fluceqns}))
there could be a host of observables associated with these domains:
anomalous distribution of pions at small transverse momentum, enhanced
production of direct photons with a distinct polarization asymmetry
(net helicity) and strong fluctuations in the particle multiplicity on
an  event by event basis~\cite{dccspino}. All of these are potentially
important phenomena if they occur during or after the QCD phase
transition in the early Universe. Because the pions are pseudoscalar
particles, if they are produced coherently through spinodal
decomposition and they eventually produce photons, this could give rise
to primordial magnetic fields with {\em net helicity} because of the
polarization asymmetry of the produced photons. This possibility,
however, must be analyzed further for a reliable assessment.

There has been a substantial experimental effort by the WA98
collaboration at SPS-CERN  and by the Minimax collaboration at the
Tevatron (Fermilab) to detect the signatures associated with DCC's, but
so far the search has yielded negative results~\cite{dccexp}.
Furthermore, recently the NA44 collaboration at SPS-CERN has reported
its experimental study of critical fluctuations on an event-by-event
basis in $Pb+Pb$ collisions. The analysis of the data does not reveal
any large fluctuations that could be associated with critical
phenomena.

These results  disappointing as they may be should not be interpreted,
yet, as the implausibility of the physical mechanism. It is conceivable
that the temperature region probed by SPS is not high enough to lead to
a strongly out of equilibrium chiral transition, which would be needed
for the above mechanism to be viable~\cite{dccspino}.

Thus the study of critical fluctuations and the possibility of large
pion domains will be continued at RHIC.

\subsection{Little bang vs. Big Bang:}

\begin{figure}[h]
\begin{center}
\epsfig{file=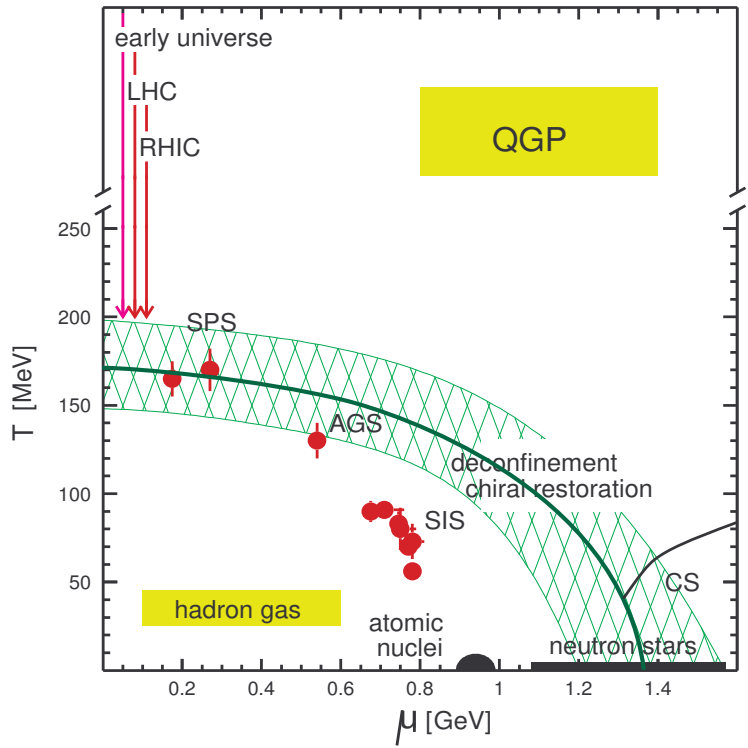,height=4in,width=4in}
\caption{QGP phase diagram. Adapted from~\cite{braun}}
\label{fig:qgpphase}
\end{center}
\end{figure}

Having reviewed the standard Big Bang cosmology and the experimental
effort to study the QCD phase transition(s) and to map the QCD phase
diagram with accelerators, we now compare the settings for the QCD
phase transition during the Big Bang to that prevailing in accelerator
experiments, i.e, the ``little bang''. The QCD phase diagram and the
different regions studied with accelerator experiments as well as the
region of temperatures and chemical potential prevailing during the
first $\mu$sec after the Big Bang is depicted in figure
(\ref{fig:qgpphase}) (for details on the CS (color superconducting)
phase(s) see ~\cite{raja}). For a thorough review of the experimental
aspects to study the phase diagram see~\cite{braun}.


There are important similarities as well as differences between the
situations in which the QCD phase transition occurs in the early
Universe and in accelerators which are worth summarizing.

\begin{figure}[h]
\begin{center}
\epsfig{file=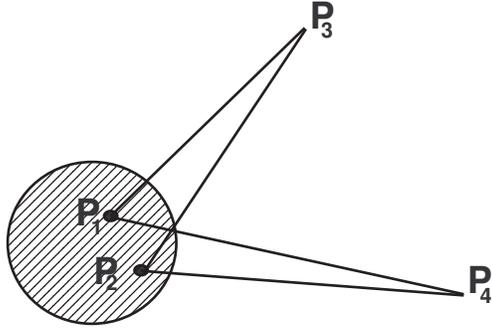}
\caption{HBT interferometry.} \label{fig:hbt}
\end{center}
\end{figure}

\begin{itemize}
  \item {\bf Space-time scales:} The space-time scales are very
  different: in the early Universe the QCD phase transition occured at
  $t\sim 1-10 \times 10^{-6}~{\rm secs}$ after the Big Bang when the
  size of the horizon was $d_H \sim 10 ~{\rm Km}$ the expansion rate was
  $H \sim 10^6~ secs^{-1}$. Quark and
  gluon mean free paths are of order $\sim 1~ {\rm fm}$, hence conditions for LTE prevailed.

   In the present Universe, i.e,  in heavy ion collision experiments,
   the lifetime of the QGP is estimated to be $\sim 10 ~ fm/c$ and the
   typical size is $\sim 10 fm$, the expansion time scale is $\sim~ 1 c/fm$ while
   the mean free paths are still  of order  $\sim 1 fm$. Therefore departures
   from LTE and
   non-equilibrium effects could play an important role in heavy ion
   collisions, while in the early Universe LTE     is a very good approximation.
  \item {\bf Baryon density:} In the early Universe, the QCD phase
  transition ocurred in an almost baryon free environment. This is
  because the entropy was dominated by photons (and neutrinos) with a
  baryon to entropy ratio $n_B/s \sim 10^{-9}$. At RHIC and LHC
  energies it is expected that the central collision (central rapidity)
  region will be almost baryon free, with most of the baryons in the
  fragmentation regions and the entropy dominated by (almost massless)
  pions with a very small ratio $n_B/n_{\pi}$ in the central region.
  \item {\bf LSS,CMB...freeze out,HBT} There is a direct analogy between
  photon decoupling, the last scattering surface (LSS) in cosmology and the
  freeze-out of hadrons: in the case of photon decoupling, the mean
  free path for Thompson scattering becomes of the same order as the
  Hubble radius and the photons consequently free-stream. A similar
  situation occurs for hadrons, when their mean free path becomes
larger than the typical separation of hadrons, there is no further
rescattering and the hadrons free stream towards the detector. Just as
in the case of the CMB where correlations in the temperature
anisotropies averaged over the sky give information on the last
scattering surface, there is a similar technique for hadrons, mainly
pions, which makes use of the Hanbury-Brown-Twiss (HBT) interferometric
effect~\cite{hbt,hbt2,meyer}. Consider a source that emits {\em
identical} particles from positions $P_1$ and $P_2$ and these particles
are later observed at points $P_3$ and $P_4$ as envisaged in fig.
(\ref{fig:hbt}). Because of the symmetrization of the pion
wavefunctions, both emission points contribute to the observable at
both reception points, even if the particles are non-interacting. The
correlations in the momenta of the two pions are studied by defining
\begin{equation}\label{hbtcorr}
C({\vec p},{\vec q}) = \frac{{\cal P}_2({\vec p},{\vec q})}{{\cal
P}_1({\vec p}){\cal P}_1({\vec q})}
\end{equation}
\noindent with ${\cal P}_2({\vec p},{\vec q})$ the joint
probability of two pions with momenta $\vec p$ and $\vec q$
respectively and the ${\cal P}_i$ are the individual
probabilities, so that if $C({\vec p},{\vec q}) =1$ the events are
uncorrelated.

The Fourier transform of ${\cal P}_2$ gives information on the
space-time structure of the source that emitted the
pions~\cite{hbt,hbt2,meyer}. Thus by studying pion interferometry in
heavy ion collisions one learns about the space-time structure of the
hadronic gas at the freeze-out surface, just as the multipole expansion
of the correlation function of the temperature anisotropies  gives
information about the cosmological parameters at the last scattering
surface~\cite{turner}. In particular HBT interferometry can be used to
study signals from a first order phase transition as well as coherent
pion production (DCC's)~\cite{hbt2} and is therefore an important
diagnostic tool for observable consequences of the QCD phase
transition(s), just as the analysis of the temperature anisotropies in
the CMB.
\end{itemize}


\subsection{QGP in the core of pulsars:}

Although I  have focused the discussion of observable consequences
of the QCD phase transition on the experimental signatures at SPS
and RHIC, corresponding to a {\em hot} QGP,  in section 3.7 I have
presented arguments  suggesting that the
 core of neutron stars may have a deconfined phase of quarks and
 gluons. In neutron stars the typical temperatures are of order $1 ~{\rm Mev}$ while the
 baryochemical potential is of order $\sim {\rm Gev}$. Therefore a deconfined phase of quarks
 and gluons in this case correspons to a cold degenerate QGP~\cite{glen,weber,glenweber}.

 In a non-rotating neutron star, the boundary between the
 deconfined, mixed (hadron and QGP) and hadronic phases are fixed, but
 in a rotating neutron star (pulsar) these boundaries change as the
 rotational frequency of the star changes in time. In ref.~\cite{pei}
 it was pointed out that since the compressibility of normal nuclear
 matter phase and the deconfined (almost free Fermi gas) of the QGP are
 {\em different} (the incompressibility of normal nuclear matter is greater than that
 of an almost free Fermi gas of quarks) a structural change will occur upon a change in
 frequency. This has important consequences in the spin-up or spin-down
 stages of millisecond pulsars~\cite{pei,glenweber}. As a  millisecond
 pulsar spins-down, its central density may rise above the critical
 density for the QGP phase transition in dense nuclear matter and the
 central core changes to a phase with a softer equation of state. In
 ref.~\cite{glenweber} the argument is reversed to contemplate the
 spin-up stage of a X-ray pulsar in an low-mass X ray binary (LMXB's).
 In this case the X-ray pulsar accretes mass from its binary companion
 which is typically a low mass white dwarf, and spins-up. In this case
 if the density at the core {\em falls below} the critical density the,
 QGP at the core turns into the hadronic phase and the existing quark
 matter at the core is ``spun-out''. In this case,  the frequency of
 these X-ray pulsars {\em increases} during accretion~\cite{glenweber}.
 The authors of ref.~\cite{glenweber} argue that for a range of frequencies  the changes in the
 quark matter composition at the core will {\em inhibit} changes in the rotation
 frequency of the pulsar  because of the increase in the moment of
 inertia. The net result of ref.~\cite{glenweber} is that these
 accreters will spend more time near these critical frequencies
 resulting in an anomalous distribution of frequencies at or near this
 frequency. A recent analysis of the oscillations of millisecond
 pulsars by the Rossi X-ray Timing Explorer (RXTE)~\cite{klis} clearly shows
 that there is a frequency, $\sim 300$ Hz at which the pulsar
 distribution peaks. A detailed study of the time evolution of the
 moment of inertia and the rotational frequency for an LMXB was
 performed in ref.~\cite{glenweber}. For the case in which the
 core of the pulsar has a deconfined QGP phase the authors show that
 the pulsar distribution has a spike at a frequency which is compatible
 with the RXTE data. The observed and theoretical distributions in
 frequency are shown in the figure (\ref{fig:qgpns}) below.


\begin{figure}[h]
\begin{center}
\epsfig{file=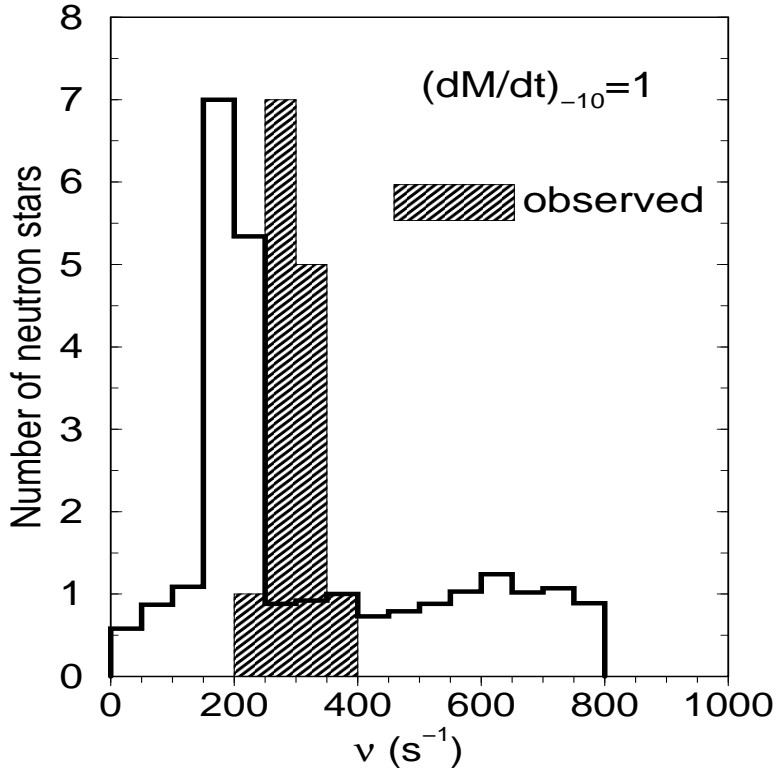,height=4in,width=4in}
\caption{Frequency distribution of X-ray neutron stars. The spike in
the calculated distribution is associated with the spinout of the quark
matter phase and the corresponding growth of the moment of inertia.
From reference~\cite{glenweber} } \label{fig:qgpns}
\end{center}
\end{figure}


 Glendenning and Weber~\cite{glenweber} argue
that a degenerate quark matter core (cold and dense QGP) in the
pulsars of some  LMXB's of suitable masses can resist spin-up
through the ongoing reduction of the quark matter cores in these
accreting pulsars. As explained by these authors, a conversion
from the {\em quark to hadronic} matter (the inverse situation as
envisaged from a cooling QGP) manifests itself in an expansion of
the star and a significant increase in the moment of inertia. The
angular momentum added to the pulsar via accretion is used up by
the star's expansion, inhibiting the spin-up until all the quark
matter in the core has been transformed into the mixed (or
hadronic) phase.

These authors conclude  that while there are several possibilities to
explain the spike in the frequency distribution, this mechanism can
contribute to explaining the anomalous frequency distribution.

Other potential signals of a quark-hadron phase transition in
neutron stars had been suggested, such as changes in the surface
temperature~\cite{tempns} as well as rotational mode
instabilities~\cite{rotns}. This wealth of potential observables
can provide a definite {\em astrophysical} evidence of a
deconfined phase of quarks and gluons in some of the most extreme
environments in the {\em present} Universe.

As mentioned above, there are fascinating novel color
superconducting phases that are conjectured to arise in cold and
dense QCD. The observational aspects of these novel phases of QCD
are still being investigated and we refer the reader to the review
articles in the literature (\cite{raja}) for   details.

\section{Back to the early Universe: summary}

The main goal of these lectures is to assess the observational
possibilities of phase transitions in early Universe cosmology.  I
argued that the analysis of CMB anisotropies can reveal
information on phase transitions that occurred during inflation.

Phase transitions during the inflationary epoch modify the power
spectrum of the primordial density fluctuations whose wavelengths
cross the horizon during inflation and re-enter after
recombination. Current theoretical models suggest that such phase
transition occurred at a grand unified energy scale or perhaps
within some supersymmetric theory but certainly beyond the current
standard model and within a realm that neither theory nor
experiment is on solid grounds.

I have then narrowed the discussion down to the phase transitions
predicted by the standard model of particle physics because this
model is on solid theoretical and experimental footing. However
the observational consequences of phase transitions are argued to
be rather indirect.

The standard model predicts {\em two} phase transitions, one at
the electroweak (EW) scale and the other at the QCD scale. The
latter can actually be {\em two} phase transitions: a
confinement-deconfinement PT between an almost free gas of quarks
and gluons  and hadrons and the other the chiral phase transition,
both at a temperature $T \sim 150-200~{\rm Mev}$. While originally
it was conjectured that baryogenesis could be explained by a
strong first order phase transition at the EW scale, current
bounds from LEP for the Higgs mass seem to rule out a strong first
order phase transition. Furthermore, the magnitude of the CP
violating parameters in the standard model, contained in the phase
of the CKM matrix seem to be too small to lead to the observed
baryon asymmetry.  Thus, within the standard model the {\em only}
phase transition that could lead to observable consequences are
those of QCD. The QCD phase transition(s) occurred at a time $t
\sim 10^{-5}-10^{-6}$ seconds after the Big Bang when the size of
the horizon is $d_H \sim 10$ Km and the mass contained within the
horizon is $\sim 1~M_{\odot}$. There are several possible
consequences of the QCD phase transition that have been reviewed:
i) Inhomogeneous nucleosynthesis, which could lead to a possible
explanation of the new constraints for $\Omega_bh^2$ from the
Boomerang/Maxima data analysis, solar mass primordial black holes,
that could be part of the cold dark matter and strange quark
nuggets or strangelets that can also be a component of CDM. All of
these proposals are based on a first order phase transition and
the details require a deep knowledge of the QCD parameters and
equation of state. The lattice gauge theory program is providing
reliable data on these issues but clearly there is a  need for an
experimental program to understand the feasibility and reliability
of these observables. The experimental study of the EW phase
transition itself is not feasible within our lifetime (certainly
mine!) since energy densities ten orders of magnitude larger than
that in nuclear matter are required. However the QCD phase
transition(s) require energy densities a few times that of nuclear
matter and are currently being studied by accelerators at CERN
(SPS) and BNL (RHIC) with a forthcoming upgrade at CERN(LHC).

Furthermore, after an excursion into the lives and deaths of stars
I presented recent results that suggest that a deconfined phase of
quarks can exist at the cores of neutron stars.

I then summarized the program of relativistic heavy ion collisions
that seeks to map the phase diagram of QCD and provided the
recently reported analysis of ``evidence'' gathered at the
SPS-CERN. While the interpretation of this evidence does not
uniquely point to the discovery of a novel form of matter: the
quark-gluon plasma (QGP), thus proving the confinement-deconfinement
phase transition in QCD, taken together they represent a
formidable body of ``circumstantial evidence'' in its favor.

From the experimental data on $J/\Psi$ suppression as well as photons
and dileptons we will learn about the physics of the deconfined hot
quark gluon plasma phase as well as  the hot hadron phase. From data on
strangeness production we can learn about the formation of strangelets
during the QCD phase transition in the early Universe. From flow we can
learn the Eos of QCD and if coherent pion domains are formed and
measured we can then provide a more sound assessment on the possibility
of primordial magnetic fields with net helicity seeded by the decay of
these coherent pion domains. Thus the experimental program will
undoubtedly lead to a firmer physical picture and a more solid basis
for theoretical work.

Furthermore, recent astrophysical observations of the frequency
distribution of pulsars, which show an anomalous spike at $\sim
300$ Hz {\em could} in fact already be a telltale signal of a
novel phase of cold and dense QCD in which deconfined quarks and
gluons make up the core of pulsars.

After SPS shuts down paving the way to the LHC, the torch is
passed on to the Relativistic Heavy Ion Collider at BNL and the
future LHC at CERN in which the ALICE program will study heavy ion
reactions. Experiments at these colliders have the potential for
understanding the details of the QCD phase transitions. Current
and future astrophysical observations of X-ray spectra, timing and
rotational properties of pulsars could confirm the possibility of
deconfined quark matter at the core of these compact stars. Thus,
indeed RHIC, LHC and pulsars are opening a window to the early
Universe and giving a glimpse of its infancy when it was only
$10^{-6}$ seconds young.

{\bf Acknowledgements:}  The author thanks NSF for support through
grants PHY-9605186, PHY-9988720 and  NSF-INT-9815064.

\end{document}